\documentclass[preprint,12pt]{elsarticle}

\usepackage{amssymb}
 \usepackage{amsthm}

\usepackage{hyperref}
\usepackage[table]{xcolor}
\usepackage{enumerate}
\usepackage{amsmath}
\usepackage{float}
\usepackage{amssymb}
\usepackage{graphicx}
\usepackage[latin1]{inputenc}
\usepackage{amsfonts}

\usepackage[nodots,nocompress]{numcompress}

\journal{Ship Technology Research}

\begin{document}

\begin{frontmatter}

 \author{Robinson Peri\'c\fnref{label2}}
  \author{Moustafa Abdel-Maksoud}
 \fntext[label2]{Corresponding author. Tel. +49 40 42878 6031. \\ 
  \textit{E-mail adress:} \href{mailto:robinson.peric@tuhh.de}{\nolinkurl{robinson.peric@tuhh.de}}}
 \address{Hamburg University of Technology (TUHH),
Institute for Fluid Dynamics and Ship Theory (M8), Hamburg, Germany}

\title{Reliable Damping of  Free Surface Waves in Numerical Simulations}


 

\begin{abstract}
This paper generalizes existing approaches for free-surface wave damping via momentum sinks for flow simulations based on the Navier-Stokes equations. 
It is shown in 2D flow simulations that, to obtain reliable wave damping, the coefficients in the damping functions must be adjusted to the wave parameters. 
A scaling law for selecting these damping coefficients is presented, which enables similarity of the damping in model- and full-scale. The influence of the thickness of the damping layer, the wave steepness, the mesh fineness  and the choice of the damping coefficients are examined. An efficient approach for estimating the optimal damping setup is presented.  
Results of 3D ship resistance computations show that the scaling laws apply to such simulations as well, so the damping coefficients should  be adjusted for every simulation to ensure convergence of the solution in both model and full scale.
Finally,  practical recommendations for the setup of reliable damping in flow  simulations with regular and irregular free surface waves are given.

\end{abstract}


\end{frontmatter}


\newcommand*{\factor}{1.0}
\newcommand*{\geomfactor}{0.8}
\newcommand*{\halffactor}{0.495}

\section{Introduction}
\label{intro}
In free surface flow simulations based on the Navier-Stokes equations, it is often required to model an infinite domain, which basically means minimizing undesired wave reflections at wave-maker and domain boundaries, while choosing the solution domain as small as possible in order to lower the computational effort. 
Since such simulations are among the computationally most effortful techniques to solve numerical flow problems, they are employed when simpler approaches (e.g. potential flow methods) cannot be used or when higher accuracy is required. As the schemes contain numerical errors (discretization, iteration, modelling (in some cases)), to achieve the required accuracy it is necessary to minimize avoidable uncertainties; of these, the performance of the wave damping approach is often among the most critical.

The elimination of wave reflections at the domain boundaries is commonly achieved by 
\begin{enumerate}[(i)]
\item \textit{increasing the domain size}
\item \textit{beaches} (e.g. Lal and Elangovan (2008)): a slope in the domain bottom leads to wave breaking and energy dissipation as in experiments
\item  \textit{grid damping} (e.g. Kraskowski (2010),  Peri\'c and Abdel-Maksoud (2015)): continuously increasing the cell size towards the corresponding domain boundary increases numerical discretization and iteration errors, thus damping the wave (this approach is also called numerical beach or grid extrusion)
\item \textit{active wave absorption techniques} (e.g. Cruz (2008); Higuera et al. (2013); Sch\"affer and Klopman (2000)): a boundary-based wave-maker generates waves which eliminate the incoming waves via destructive interference
\item \textit{solution-forcing} or \textit{solver-coupling} (e.g. Ferrant et al. (2008), Guignard et al. (1999), Kim et  al. (2012), Kim et al. (2013), W\"ockner-Kluwe (2013)): the flow is forced to a known solution in the vicinity of the boundary or the Navier-Stokes-based flow solver is coupled to another (e.g. potential flow based) solver
\item \textit{damping layer approaches} (e.g. Cao et al. (1993); Choi and Yoon (2009); Ha et al. (2011); Israeli and Orszag (1981); Park et al. (1999)): the damping layer  (also called sponge layer, absorbing layer, damping zone, porous media layer) is a zone set up next to the corresponding boundaries, in which momentum sinks are included in the governing equations to damp the waves propagating through the zone
\end{enumerate}
Apart from the above methods,  further approaches have been developed for other governing equations, like potential flow with boundary element method or Boussinesq-type equations (see e.g. Grilli and Horrillo (1997) and references therein). However, many of these approaches have not yet been transferred to Navier-Stokes-type equations.
From the above mentioned approaches, (i) is the least feasible due to its (in many cases enormous) inherent increase in computational effort. With (ii) it is difficult to minimize reflections to less then $10\%$ of the incoming wave, which is a problem in experiments as well. Approaches (iv) and (v) have recently attracted much interest and produced good results; however, since they are often used to simultaneously generate and damp waves, it is likely that their damping performance varies depending on the kind of waves they are currently generating. At the time of writing, (iii) and (vi) are the most widely-used wave damping approaches in Navier-Stokes-type equation flow solvers, implemented in most commercial and research codes. Although both approaches have been used with success, the disadvantage of (iii) is that its performance depends on grid, time step, temporal/spatial discretization schemes, etc., which for sufficiently fine discretization is not the case for (vi), as the results in this work suggest. Thus (iii) is less predictable than (vi) regarding the damping quality.  This work focuses exclusively on damping layer approaches (vi) based on momentum source terms, since a detailed investigation and comparison of all previously mentioned approaches was not possible in the scope of this study.

This paper discusses momentum source terms based on linear or quadratic damping functions as defined in Sect. \ref{general}. 
The amount and character of the damping is controlled by coefficients in the damping functions and the thickness of the damping layer. 
Such approaches are widely used and already implemented in several open source as well as commercial computational fluid dynamics (CFD) solvers. Section \ref{general} shows how the existing implementations can be generalized.
Two of the most widely used implementations, the ones from CD-adapco STAR-CCM+, based on Choi and Yoon (2009), and ANSYS Fluent, based on Park et al. (1999),  are discussed in Sect. \ref{implem}.  Various other implementations of linear or quadratic damping exist, see e.g. Cao et al. (1993)  and Ha et al. (2011). 

The nomenclature from Sect. \ref{implem} is used throughout the work. 
For all simulations in this study, the damping function from Choi and Yoon (2009) is used, since it is rather generic and can be set up to act similar to the other approaches mentioned. Thus the results are easily applicable to  all damping approaches used in other CFD codes which can be generelized as described in Sect. \ref{general}.
Although widely used, much about the damping functions remains unknown at the time of writing.  
It has been observed that the coefficients in the damping functions, and thus the damping performance, are case-dependent. 
In the above mentioned codes, these damping coefficients can be modified by the user. However, no guidelines seem to exist for this. 
Thus in practice, either the default settings or values from experience  are used as coefficients in the damping functions. If during or after the simulation it is observed that the damping does not work satisfactorily, then the damping coefficients are modified by trial and error and the simulation is restarted. This procedure is repeated until acceptable damping is obtained, a process which requires human interaction and can cost  considerable additional time and computational effort. 
Especially in the light of the increased automation of CFD computations, it would be preferable to be sure about the damping quality before the simulation.
Similar to wave damping in experiments, where according to Lloyd (1989) seldom data on beach performance is published, also with CFD simulations, the performance of the wave damping is often not sufficiently accounted for. 
Thus with most CFD publications, it is difficult to judge on the damping quality even if the damping setup is given.  

The aim of the present work is to clarify how the damping functions work, on which factors the damping quality depends  and how reliable wave damping can be set up case-independently, so that no more fine-tuning of the damping coefficients is necessary. 

In Sect. \ref{general}, the generalized forms of linear and quadratic damping functions are given. In the following Sect. \ref{implem}, the damping approach used in this work is described. Moreover, it is shown exemplarily for two widely used damping functions how these can be generalized to the equations given in Sect.  \ref{general} and how results from one implementation can be transferred to another.

Starting from the analogy of the damped harmonic oscillator, Sect. \ref{Lf1} attempts to show similarities to damping phenomena outside the hydrodynamics field and constructs scaling laws for wave damping based on this information. These scaling laws are fully formulated and verified  in Sects. \ref{Scalel2f1} and \ref{Scalel2f2}. Furthermore, recommendations for selecting the damping coefficients  are given. 

Sections \ref{l2f1} to \ref{heightinfl} investigate via 2D flow simulations how the damping quality and behavior is influenced by the choice of the coefficients in the damping functions, the thickness of the damping layer, the computational grid and the wave steepness.
As the investigations so far have been mainly concerned with regular, monochromatic waves, recommendations for  the damping for irregular waves are given in Sect. \ref{irrwav} and illustrated with simulation results.

Since wave damping is also widely used to speed up convergence for example in ship resistance computations, it is verified in Sect. \ref{kcs} via 3D flow simulations that the presented scaling laws hold for such cases as well.

\section{Governing Equations and Solution Method}
\label{goveq}
The governing equations for the simulations are the Navier-Stokes equations,
which consist of the equation for mass conservation and the three equations for
momentum conservation: 
\begin{equation}
\frac{\mathrm{d}}{\mathrm{d} t} \int_{V} \rho \ \mathrm{d}V + \int_{S} \rho \textbf{v}  \cdot \textbf{n} \ \mathrm{d}S =  0 \quad ,
\label{conti}
\end{equation}
\begin{align}
\frac{\mathrm{d}}{\mathrm{d} t} \int_{V} \rho u_{i} \ \mathrm{d}V 
+ \int_{S} \rho u_{i} \textbf{v}  \cdot \textbf{n} \ \mathrm{d}S =  \nonumber \\ 
\int_{S} (\tau_{ij}\textbf{i}_{j} - p\textbf{i}_{i}) \cdot \textbf{n} \ \mathrm{d}S 
+ \int_{V} \rho \textbf{gi}_{i} \ \mathrm{d}V + \int_{V} q_{i} \ \mathrm{d}V \quad .
\label{navier_stokes}
\end{align}
Here $V $ is the control volume (CV) bounded by the closed surface $\mathrm{S}$, \textbf{v} is the velocity vector of the fluid with the Cartesian components $u_{i}$, \textbf{n} is the unit vector normal to $S$ and pointing outwards, $t$ is time, $p$ is the pressure, $\rho$ and $\mu$ are fluid density and dynamic viscosity, $\tau_{ij}$ are the components of the viscous stress tensor, \textbf{i}$_{j}$ is the unit vector in direction $ x_{j} $, \textbf{g} comprises the body forces and $ q_{i} $ is an optional  momentum source term. 
For the present simulations, the  only body force considered was the gravitational acceleration, i.e. \textbf{g}$ = (0, 0, -9.81\frac{ \mathrm{m}}{\mathrm{s}^{2}})^{T}$.
Only incompressible Newtonian fluids are considered in this study. Thus $ \tau_{ij} $ takes the form
 \[ \tau_{ij} = \mu \left( \frac{\partial u_{i}}{\partial x_{j}} + \frac{\partial u_{j}}{\partial x_{i}} \right)  \quad .\]
No turbulence modeling was applied to the above equations since, unless waves break, the flow inside them can be considered practically laminar and all structures of interest can be resolved with acceptable computational effort.
For all simulations, the software STAR-CCM+ $8.02.008$ was used.
The volume of fluid (VOF) method  implemented in the STAR-CCM+ software is used to account for the two phases (air and water). Further details on the method can be found in Muzaferija and Peri\'c (1999).
The governing equations are applied to each cell and discretized according to the Finite Volume Method (FVM). All integrals are approximated by the midpoint rule. The interpolation of variables from cell center to face center and the numerical differentiation are performed using linear shape functions, leading to approximations of second order. The integration in time is based on assumed quadratic variation of variables in time, which is also a second-order approximation. Each algebraic equation contains the unknown value from the cell center and the centers of all neighboring cells with which it shares common faces. The resulting coupled equation system is then linearized and solved by the iterative STAR-CCM+ implicit unsteady segregated solver, using an algebraic multigrid method with Gauss-Seidl relaxation scheme, V-cycles for pressure and volume fraction of water, and flexible cycles for velocity calculation.
For each time step, one iteration consists of solving the governing equations for the velocity components, the pressure-correction equation (using the SIMPLE method for collocated grids to obtain the pressure values and to correct the velocities) and the transport equation for the volume fraction of water. 
For further information on the discretization of and solvers for the governing equations, the reader is referred to  Ferziger and Peri\'c (2002) or the STAR-CCM+ software manual.
 
\section{General Formulation for Linear and Quadratic Damping}
\label{general}
Linear wave damping is obtained by inserting the momentum source term  $ q^{\rm d, lin}_{i}  $ for $ q_{i} $ in Eq. (\ref{navier_stokes}): 
\begin{equation}
q^{\rm d, lin}_{i} = \rho  C_{i,{\rm lin}} u_{i} \quad ,
\label{lin}
\end{equation}
with coefficient $ C_{i,{\rm lin}} $, which usually depends on the spatial location. $ C_{i,{\rm lin}} $ regulates the strength of the damping and is used to  provide a smooth blending-in of the damping, by increasing the amount of damping from weak damping where the waves enter the damping layer to strong damping at the end of the damping layer. This is to prevent undesired reflections at the entrance to the damping layer as shown in Sects. \ref{l2f1} and \ref{l2f2}. Commonly, the blending-in is realized in an exponential or quadratic fashion.  

Quadratic wave damping takes the form
\begin{equation}
q^{\rm d, quad}_{i} = \rho  C_{i,{\rm quad}}  |u_{i}| u_{i} \quad ,
\label{quad}
\end{equation}
with coefficient $ C_{i,{\rm quad}} $, which regulates strength and blending-in of the damping. In contrast to Eq. (\ref{lin}), fluid particles with higher velocities experience disproportionately high damping.

In most CFD codes, $ C_{i,{\rm lin}} $  and $ C_{i,{\rm quad}} $ can be adjusted by the user. Usually, but not necessarily, the momentum source terms are only applied to the equation for the vertical velocity component.

\section{Common Implementations of Linear and Quadratic Wave Damping}
\label{implem}
\begin{figure}[H]
\includegraphics[width=\factor\linewidth]{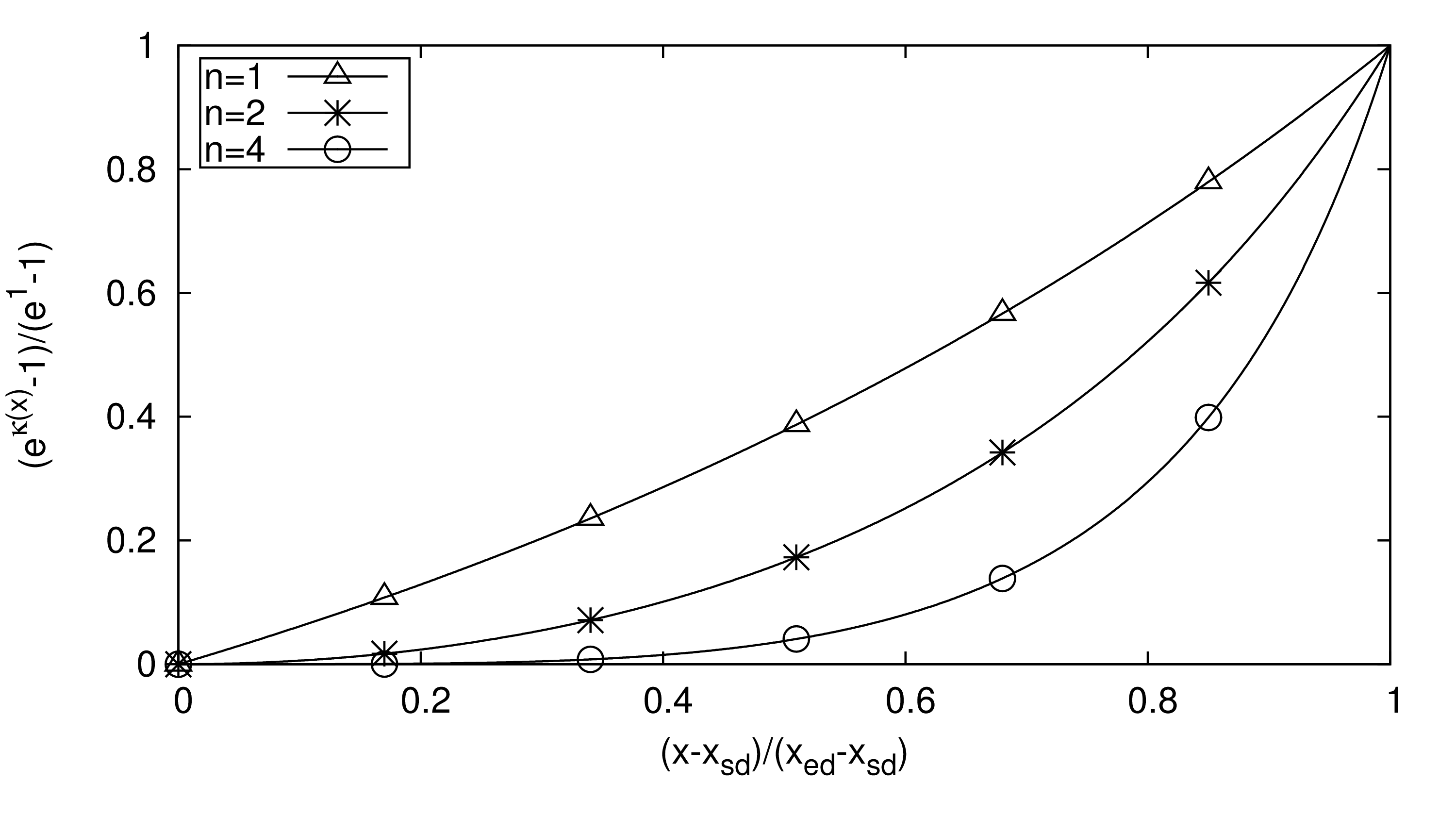}\\
\caption{The function $\frac{e^{\kappa(x)} - 1}{e^{1} - 1}$ evaluated for $n=1$, $n=2$ and $n=4$ over the dimensionless distance $\frac{x - x_{\rm sd}}{x_{\rm ed} - x_{\rm sd}}$;
a wave enters the damping layer at $\frac{x - x_{\rm sd}}{x_{\rm ed} - x_{\rm sd}}=0$ and is damped during propagation towards $\frac{x - x_{\rm sd}}{x_{\rm ed} - x_{\rm sd}}=1$, where the damping layer ends} \label{FIGkappa}
\end{figure}

A widely used approach is the one in Choi and Yoon (2009), which is e.g. implemented in  the commercial software code STAR-CCM+ by CD-adapco.
It features a combination of linear and quadratic damping, which allows the use of either one or a combination of both approaches.
Taking $ z $ as the vertical direction (i.e. perpenticular to the free surface) and $w$ as the vertical velocity component, then the following source term appears for $ q_{i} $ in Eq. (\ref{navier_stokes}):
\begin{equation}
q^{\rm d}_{z} = \rho (f_{1} + f_{2}|w|)\frac{e^{\kappa} - 1}{e^{1} - 1} w \quad ,
\label{damp1}
\end{equation}
\begin{equation}
\kappa = \left( \frac{x - x_{\rm sd}}{x_{\rm ed} - x_{\rm sd}} \right)^{n} \quad .
\label{damp2}
\end{equation}
Here $ x $ stands for the wave propagation direction with $ x_{\rm sd} $ being the start and $ x_{\rm ed} $ the end $x$-coordinate of the damping layer, thus the thickness of the damping layer $x_{{\rm d}} = |x_{\rm ed} - x_{\rm sd}|$.  $ f_{1} $ is the damping constant for the linear part and $ f_{2} $ is the damping constant for the quadratic part of the damping term. The default values are  $ f_{1} = 10.0  \, \mathrm{s^{-1}}$ and  $ f_{2} = 10.0 \, \mathrm{m^{-1}}$ according to the STAR-CCM+ manual (release $8.02.008$).
The term $ \frac{e^{\kappa} - 1}{e^{1} - 1} $ blends in the damping term, i.e. it is zero at $ x_{\rm sd} $, and it equals one at $ x_{\rm ed} $.
$ \kappa $ describes via $n$ the character of the blending functions, i.e. for $n = 1$ the blending is nearly linear. When increasing $n$, the blending becomes smoother at the entrance to the damping layer and at the same time more abrupt at $ x_{\rm ed} $, see Fig. \ref{FIGkappa}.
Another widely used approach was presented by Park et al. (1999), which is implemented in ANSYS Fluent.:
\begin{equation}
q^{\rm d}_{z}  =  \rho (0.5 f_{3}|w|)\kappa\left(1-\frac{z-z_{{\rm fs}}}{z_{{\rm b}}-z_{{\rm fs}}}\right)w \quad ,
\label{dampANSYS}
\end{equation}
with damping constant $f_{3}$, $ \kappa $ as given in Eq. (\ref{damp2}) with $ n = 2 $, vertical coordinate $z$ and $z$-coordinates of the domain bottom $z_{{\rm b}}$ and the free water surface $z_{{\rm fs}}$. The default value for $f_{3}$ is $10.0 \, \mathrm{m^{-1}}$.  

This approach can be generalized to a quadratic damping function according to Eq. (\ref{quad}). It corresponds to the quadratic part of Eq. (\ref{damp1}), except for a quadratic instead of exponential blending in $x$-direction and an added vertical fade-in. The latter term is a linear fade between domain bottom, where no damping is applied, to free surface level, where full damping is applied; for practical deep water conditions of several wavelengths water depth, the influence of the $z$-fading term on the applied damping becomes small, since most wave energy is concentrated in the vicinity of the free surface.
Thus for such cases, this damping approach can be modeled using Eqs. (\ref{damp1}) and (\ref{damp2}) with accordingly adjusted coefficients. In a similar manner,  also the other damping layer approaches from Sect. \ref{intro} (vi) can be modeled. Therefore in the following, only the approach by Choi and Yoon (2009) will be considered to study wave damping. Thus the findings in this work can easily be transferred to other damping approaches.

\section{Dependence of Damping Coefficients}
\label{Lf1}
 For gravity waves, all scaling laws should be consistent with Froude scaling.
 Thus to obtain similar wave damping for waves of different scale, the blending part of $ C_{i,\mathrm{lin}} $ and $ C_{i,\mathrm{quad}} $ from Eqs. (\ref{lin}) and  (\ref{quad}) must be geometrically similar. Therefore, the thickness $x_{\mathrm{d}}$ of the damping layer must be directly proportional to the wavelength $\lambda$.
For the remaining part of  $ C_{i,\mathrm{lin}} $ and $ C_{i,\mathrm{quad}} $, which regulates the magnitude of the damping, the scaling laws can be obtained by dimensional analysis. Due to its dimension of $ [s^{-1}] $, $ C_{i,\mathrm{lin}}  $ is directly proportional to the wave frequency $ \omega $, whereas  $ C_{i,\mathrm{quad}} $ has the dimension $ [m^{-1}] $ and is thus directly proportional to $ \lambda^{-1} $ (or $ \omega^{2} $ in deep water).
Thus to achieve similar damping when changing scale and/or wave, it is necessary to: \vspace{0.2cm}\\
\fbox {
    \parbox{\linewidth}{
\textbf{Scaling Laws}
\begin{enumerate}
\item Set the damping thickness $x_{{\rm d}}$ so that it is geometrically similar to the damping layer thickness $x_{{\rm d, ref}}$ of the reference case (i.e. scale $ x_{{\rm d}} $ with the percentual change in wavelength)
\begin{equation}
x_{{\rm d}} = x_{{\rm d, ref}} \cdot \frac{\lambda}{\lambda_{{\rm ref}}} \quad ,
\label{xdscale}
\end{equation}
\item For \textit{linear damping}, scale $ f_{1} $ with the change of wave frequency $ \omega $
\begin{equation}
f_{1} = f_{1, {\rm ref}} \cdot \frac{\omega}{\omega_{{\rm ref}}} \quad ,
\label{omegascale}
\end{equation}
with wave frequency $ \omega_{{\rm ref}} $ of the reference case. This holds for all damping formulations that can be generalized by Eq. (\ref{lin}).
\item For \textit{quadratic damping}, scale $ f_{2} $ with the change of wavelength $ \lambda $ to the power of minus one
\begin{equation}
f_{2} = f_{2, {\rm ref}} \cdot \frac{\lambda_{\mathrm{ref}}}{\lambda} \quad ,
\label{omegascale2}
\end{equation}
with wavelength $\lambda_{{\rm ref}} $ of the reference case. This holds for all damping formulations that can be generalized by Eq. (\ref{quad}).
\end{enumerate} 
    }
}

\section{Numerical Simulation Setup}
\label{simsetup}
\begin{figure}[H]
\begin{center}
\includegraphics[width=\geomfactor\linewidth]{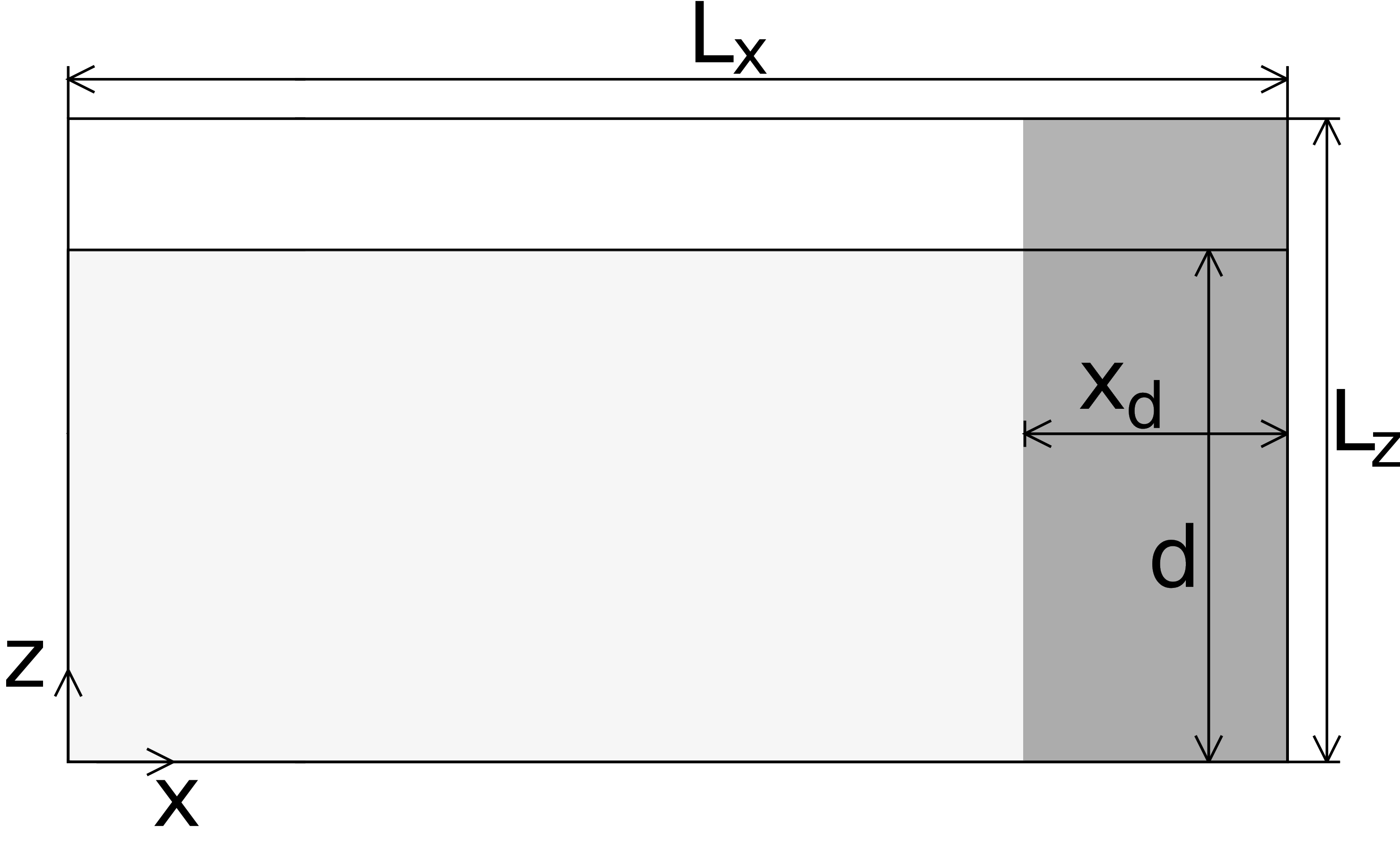}
\end{center}
\caption{2D wave tank filled with water (light gray) and air (white). The damping layer is shown in dark grey} \label{FIG2Dgridstudydomain}
\end{figure}

\begin{figure}[H]
\includegraphics[width=\factor\linewidth]{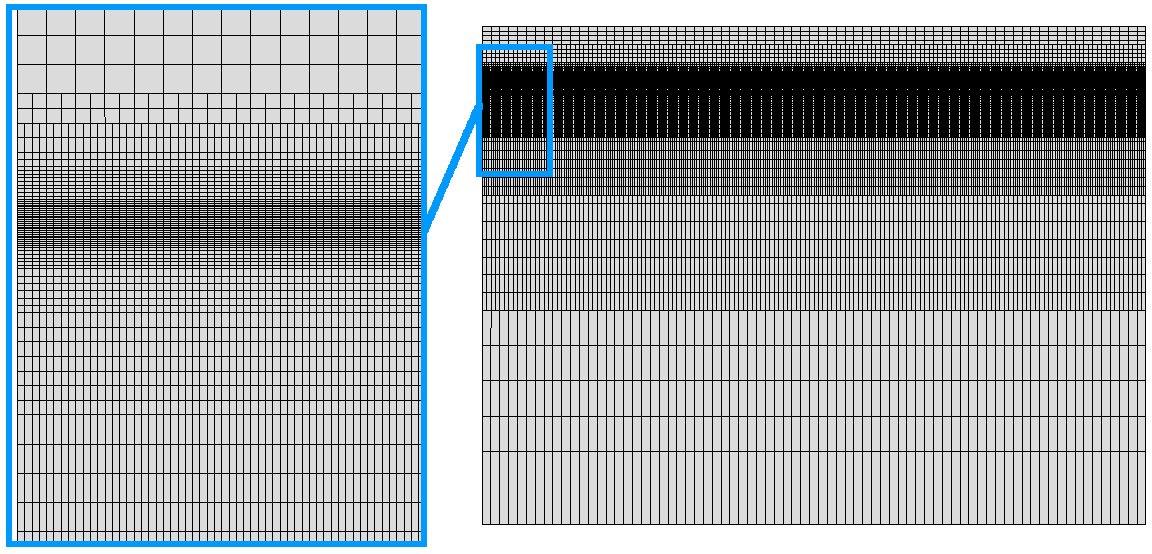}
\caption{Computational grid for the whole domain (right),  and a close up of the discretization around free surface level (left)} \label{FIGgridtsstudyGRID}
\end{figure}

Figure \ref{FIG2Dgridstudydomain} shows the solution domain, a 2D deep water wave tank with length $ L_{x} = 6\ \lambda$ and height  $ L_{z}=4.5\ \lambda $, given in relation to wavelength $ \lambda $.
It is filled with water to a depth of $ d = 4\  \lambda$, the rest of the tank is filled with air. The origin of the coordinate system is set in the bottom left front corner of the tank in Fig. \ref{FIG2Dgridstudydomain}.
The top boundary (i.e. $ z = L_{z} $) is a pressure outlet boundary, i.e. the pressure there is set constant and equal to atmospheric pressure. This corresponds to an open water tank, where air can flow in and out through the tank top. 
The  $ x =  L_{x} $ boundary is a pressure outlet, where volume fraction and hydrostatic pressure are prescribed for an undisturbed free surface.
The waves are generated by prescribing velocities and volume fraction of a $5^{{\rm th}}$-order Stokes waves according to Fenton (1985)  at the $ x = 0 $ boundary. 
The wave damping layer extends over a distance of $ x_{{\rm d}} = 2\ \lambda $ in boundary-normal direction from the  $ x =  L_{x} $ boundary.  
Thus the waves are created at  $ x = 0 $, propagate in $x$-direction, enter the damping layer at $ x = L_{x}-x_{{\rm d}} $, are subjected to damping until $ x = L_{x} $ is reached, where the remaining waves are reflected and, while further subjected to damping, propagate back to  $ x = L_{x}-x_{{\rm d}} $. If the damping was set up correctly, then the reflected waves are either completely damped or their height is decreased so much, that their influence on the simulation results can be neglected. Otherwise they will be evident in the simulation results as additional error, which can be substantial as will be shown in the following sections.
Local mesh refinement is used, so that the grid is finest in the vicinity of the free surface, where the waves are discretized by roughly $ 100 $ cells per wavelength and $ 16 $ cells per wave height, as  seen in Fig. \ref{FIGgridtsstudyGRID}. 
The temporal discretization  involved more than $500$ time steps $ \Delta t $ per wave period $ T $, thus in all investigated cases the Courant number $ C = \left( u_{i} \Delta t \right) / \Delta x_{i}$ remains well below $ 0.5 $ for every cell with size $ \Delta x_{i} $ and corresponding fluid velocity $u_{i}$.
Waves were generated for over $12$ wave periods with $8$ iterations per time step and under-relaxation of $0.4$ for pressure and $0.9$ for all other variables. 
The discretization is based on the grid convergence study conducted by Peri\'c (2013). 

The numerical setup is the same for all simulations in this study, only that the scale, wave and damping parameters are varied. Variations of the setup are mentioned where they occur. 

\section{Assessing the Damping Performance for Regular, Monochromatic Waves}
\label{assdamp}
When wave reflections occur within a damping zone, the waves will propagate back into the solution domain with a diminished height. Due to the resulting superposition of the partly reflected wave and  the original wave,  the original wave will seem to grow and shrink in a time-periodic fashion. For  regular, monochromatic waves, this occurs uniformly in all parts of the domain where the two wave systems are superposed as seen later e.g. in Sect. \ref{lengthxd} from Fig.  \ref{FIGdamplf1surfelev}. In the present work, this phenomenon is called a \textit{partial standing wave}, since the more the damping deviates from the optimal value, the more does the phenomenon resemble a standing wave, until for the bounding cases (i.e. zero damping or infinitely large damping) there is  $ 100\% $ reflection, and a standing wave occurs.

Assessing the amount of reflections for regular monochromatic waves is an intricate issue. Since only a single wave period occurs, a wave spectrum cannot be constructed.  As seen from Fig.  \ref{FIGdamplf1surfelev}, the wave height envelope changes over space. However, this cannot be predicted since it is not known in advance at which point in the damping layer the reflection mainly occurs. Thus from the recordings of a single wave probe, the height of the reflected wave cannot be determined.
However, the surface elevation in close vicinity to the boundary to which the damping zone is attached shows how much of the incident wave reaches the domain boundary despite the damping, since a maximum amplification occurs at the domain boundary when the wave is reflected.

Thus in this study, first the free surface elevation is recorded at $x = 5.75 \lambda$, i.e. next to the domain boundary. The average wave height at this location is obtained and plottet in relation to the average wave height of the undamped wave, to show how much the wave height is reduced after propagating from the entrance to the outer boundary of the damping layer.
As shown in Sects. \ref{l2f1} and  \ref{l2f2}, this is a good criterion for insufficient to optimal wave damping, where wave reflections occur mainly at the outer boundary of the damping layer, at  $ x = L_{x} $; however, this criterion will not detect reflections if the damping is too strong, since then the waves will be reflected at the entrance to the damping layer.

Therefore, secondly a reflection coefficient $C_{\mathrm{R}}$ is computed as proposed by  Ursell et al. (1960).  $C_{\mathrm{R}}$ corresponds to the ratio of the  heights of the incident wave to the reflected wave. It can be written as
\begin{equation}
C_{\mathrm{R}} = \left(H_{\mathrm{max}} - H_{\mathrm{min}}\right)/\left(H_{\mathrm{max}} + H_{\mathrm{min}}\right) \quad , 
\label{FIGcr}
\end{equation}
where $ H_{\mathrm{max}} $ is the maximum and $ H_{\mathrm{min}} $ the minimum value of the wave height envelope. 
It holds $ 0 \leq C_{\mathrm{R}} \leq 1 $, so that $C_{\mathrm{R}} = 1$ for perfect wave reflection and $C_{\mathrm{R}} = 0$ for no wave reflection.

As explained above, the wave height for a partial standing wave is not constant, but oscillates around a mean value. This occurs uniformly in all domain parts where the partial standing wave has fully developed.
Thus $ H_{\mathrm{max}} $ and $ H_{\mathrm{min}} $  are obtained in the following fashion in this work:
The free surface elevation in the whole tank is recorded at $40$ evenly timed instances per wave period starting at $10$ periods simulation time for $2$ wave periods. For each recording $j$, the average wave height $\bar{H}_{j}$ is calculated for the interval $ 2\lambda \leq x \leq 4\lambda $, which is adjacent to the damping zone but not subject to wave damping and sufficiently away from the inlet; furthermore, the partial standing wave at this location is fully developed during this time interval, while wave re-reflections at the wave-maker have not yet developed significantly. 
Therefore, the maximum  $\bar{H}_{\mathrm{max}}$ and minimum $\bar{H}_{\mathrm{min}}$ of all $\bar{H}_{j}$ values can be taken as  $ H_{\mathrm{max}} $ and $ H_{\mathrm{min}} $, respectively. 
This approach detects all wave reflections that propagate back into the solution domain. The accuracy of the scheme can be increased if the surface elevation in the tank is recorded in smaller time intervals, however this also increases the computational effort significantly. Furthermore, the above procedure requires undisturbed regular monochromatic waves. Yet in many applications, fluid-structure interactions create flow disturbances, so the above scheme cannot be applied directly in the simulations, but an additional 2D simulation of undisturbed wave propagation and damping for otherwise the same setup is required to obtain $C_{\mathrm{R}}$. As long as this process is not fully automized, running and post-processing the 2D simulation requires additional human effort.
Overall, this procedure is rather effortful, which explains why $C_{\mathrm{R}}$ is rarely computed in practice.

With the two presented approaches, it is possible to distinguish between those wave reflections, which occur mainly at the entrance to the damping zone, and those reflections which occur mostly at the boundary to which the damping zone is attached. Furthermore, the influence of wave reflections on the solution can be quantified.

\section{Variation of Damping Coefficient for Linear Damping}
\label{l2f1}
\begin{figure}[H]
\includegraphics[width=\factor\linewidth]{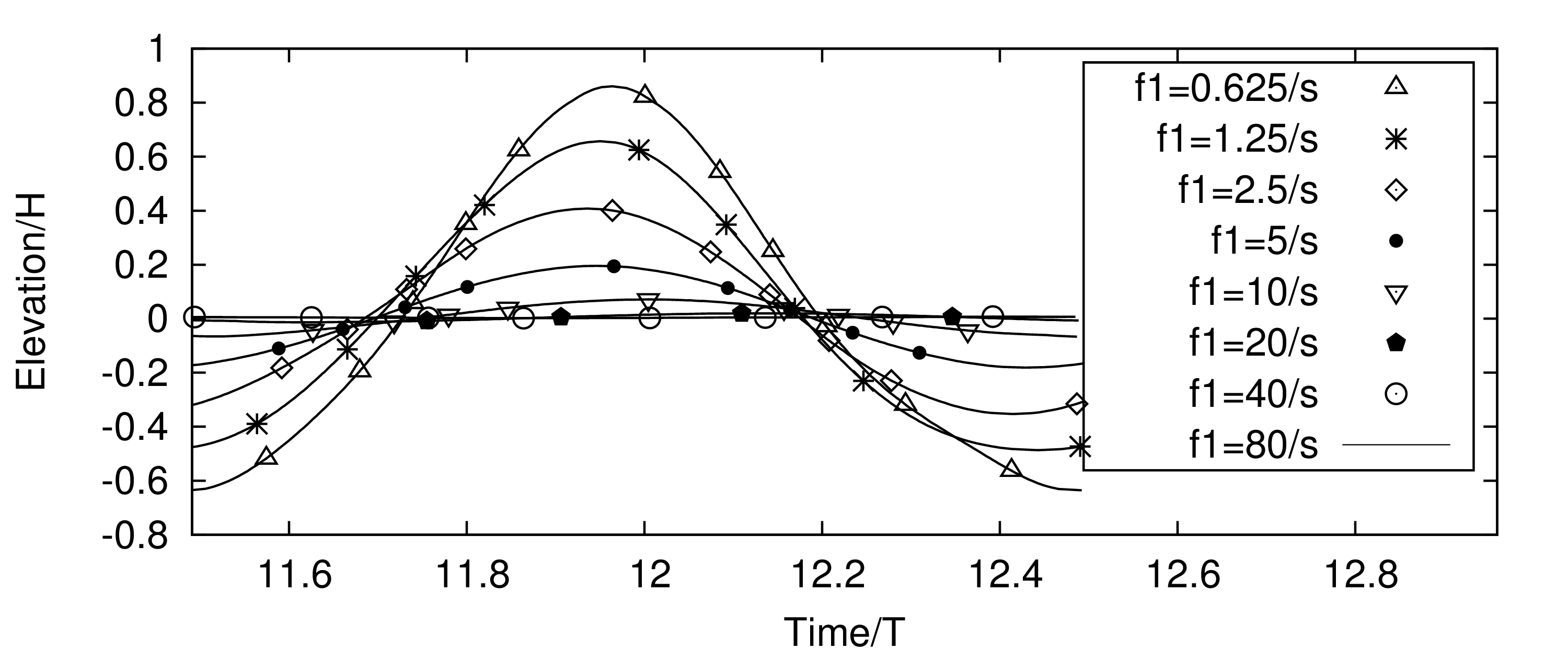}
\caption{Surface elevation scaled by height $H$ of the undamped wave over time scaled by the wave period $ T $;  recorded at $x =5.75\lambda$ for simulations with $ f_{1} \in [0.625 \, \mathrm{s^{-1}}, 80 \, \mathrm{s^{-1}}] $ and  $ f_{2} = 0$ } \label{FIGdampl2f1}
\end{figure}

To investigate which range for the linear damping coefficient according to Eq. (\ref{damp1}) produces satisfactory wave damping, waves with wavelength $ \lambda = 4\, \mathrm{m} $ and height $ 0.16 \, \mathrm{m} $ are investigated. Only linear damping is considered, so $ f_{2} = 0 $. Simulations are performed for damping coefficient $ f_{1} $  between $ f_{1} \in [0.625 \, \mathrm{s^{-1}}, 1000 \, \mathrm{s^{-1}}] $. 
The recorded surface elevations near the end of the damping layer in Fig. \ref{FIGdampl2f1} show that the damping is strongly influenced by the choice of $ f_{1} $, while the wave phase and period remain nearly unchanged.

\begin{figure}[H]
\includegraphics[width=\factor\linewidth]{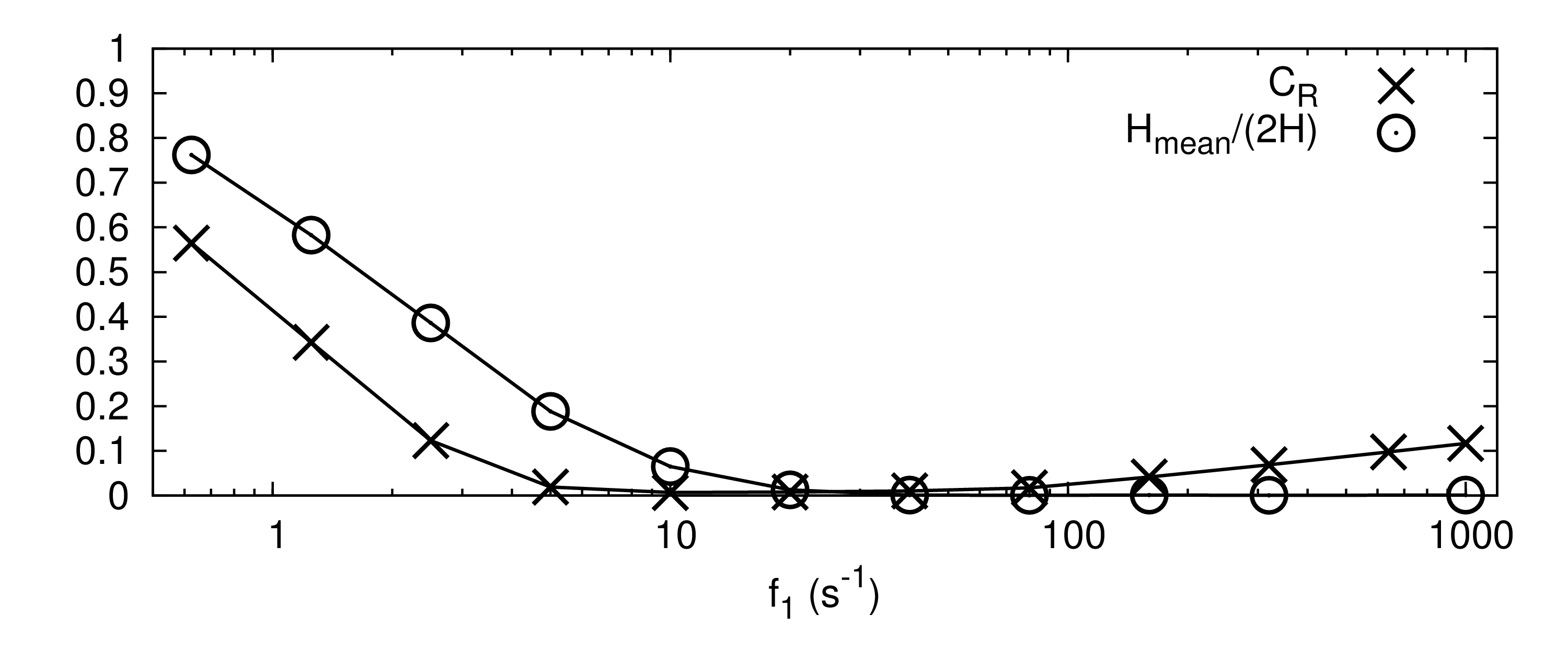}
\caption{Mean wave height $ H_{{\rm mean}} $ recorded at $x = 5.75\lambda$ scaled by twice the height $H$ of the undamped wave and reflection coefficient $ C_{\mathrm{R}} $  over damping coefficient $ f_{1} $, while $ f_{2} = 0 $} \label{FIGdampl2f1meanH}
\end{figure}

Figure  \ref{FIGdampl2f1meanH} shows how a variation of $ f_{1} $ affects the reflection coefficient $ C_{\mathrm{R}} $, which describes the amount of reflected waves that are present in the solution domain outside of the damping zone, and the mean measured wave height $H_{{\rm mean}} $ recorded in close vicinity to the boundary to which the damping layer is attached.  
\begin{itemize}
\item For smaller damping coefficients  ($ f_{1} \lesssim 20 \, \mathrm{s^{-1}} $), wave reflections occur mainly at the domain boundary to which the damping zone is attached. Here, the both show the same trend, with  $ C_{\mathrm{R}} <  H_{{\rm mean}}/(2H) $ since the reflected waves loose further height until they leave the damping zone at $x =4\lambda$. 
For $ f_{1}<1.25 \, \mathrm{s^{-1}} $, the influence of the damping is so small that a partial standing wave (as described in Sect. \ref{assdamp}) higher than the initial wave appears even inside the damping layer. 
\item For stronger damping ($ f_{1} > 20  \, \mathrm{s^{-1}}$), wave reflection occurs mainly at the entrance to the damping zone. Thus $H_{{\rm mean}} $ continually decreases when increasing  $ f_{1}$, since less of the incoming wave can pass through the damping zone, so the water surface will be virtually flat near the domain boundary. This is also visible later  in Figs. \ref{FIGdampScalel2f1freesurf_unscaled} from Sect. \ref{Scalel2f1}. The aforementioned increase in the amount of wave reflections detectable in the solution domain can be seen in the curve for  $ C_{\mathrm{R}}$. This shows that, for a given fade-in function and damping layer thickness, there is an optimum for $f_{1}$ so that the wave reflections propagating back into the solution domain are minimized. 
The best damping was achieved for  $f_{1} =10 \, \mathrm{s^{-1}}$, in which case the effects of wave reflections on the wave height within the solution domain are less than $ 0.7\% $.
\end{itemize}
This provides the following conclusions:
$H_{{\rm mean}} $ is not a suitable indicator for damping quality if the damping is stronger than optimal, however it is useful to characterize where the reflections originate from: if $ C_{\mathrm{R}} $ shows that noticeable reflections are present within the solution domain, while at the same time $H_{{\rm mean}} $ is negligibly small, then the reflections cannot occur at the domain boundary, but must occur closer to the entrance to the damping layer.
Significant reflections ($ C_{\mathrm{R}}>2\%$) in form of partial standing waves appear for roughly $ f_{1}< 5 \, \mathrm{s^{-1}} $ and $ f_{1}> 80 \, \mathrm{s^{-1}} $. The height of the partial standing wave increases the more $f_{1}$ deviates from the regime where satisfactory damping is observed; however, the increase is slower if the damping is stronger than optimal instead of weaker.

\section{Variation of Damping Coefficient for Quadratic Damping}
\label{l2f2}
\begin{figure}[H]
\includegraphics[width=\factor\linewidth]{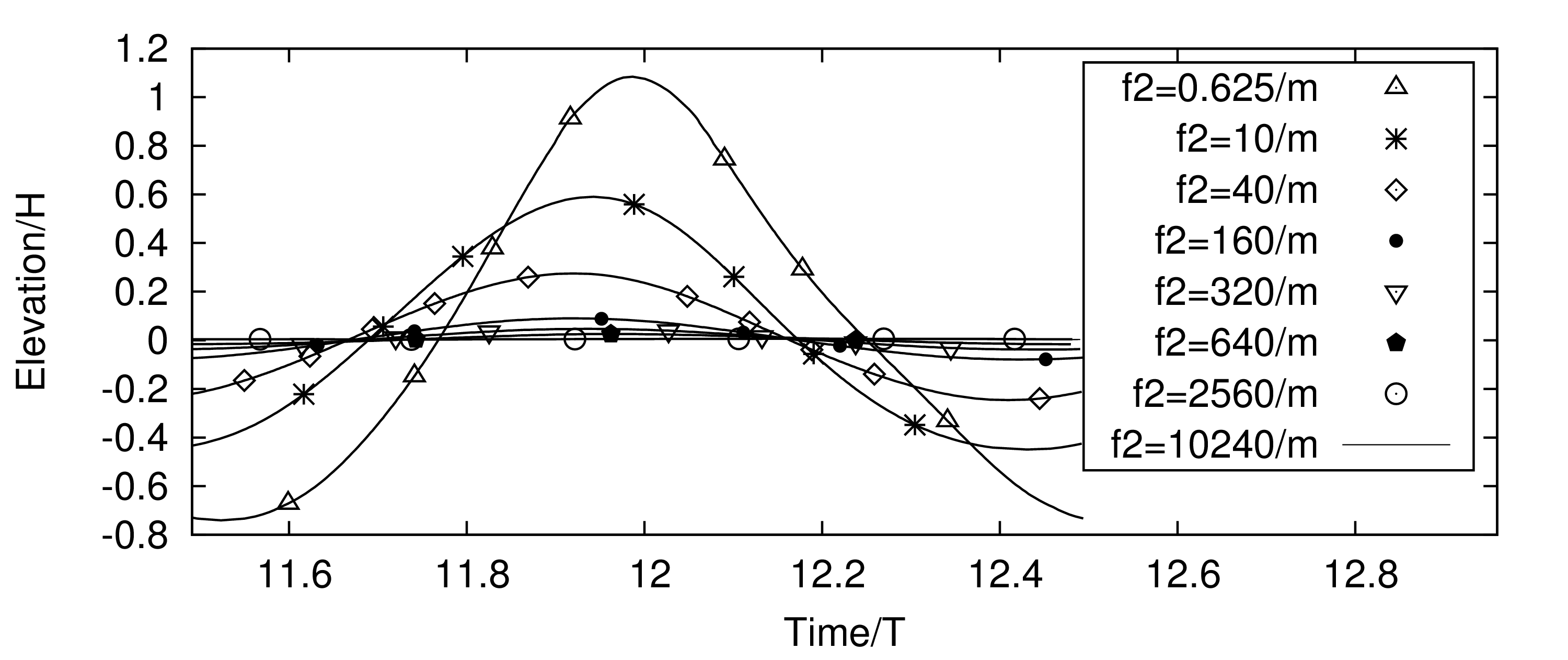}
\caption{Surface elevation  scaled by height $H$ of the undamped wave over time scaled by the wave period $ T $; recorded at $x = 5.75\lambda$ for simulations with $ f_{2} \in [0.625 \, \mathrm{m^{-1}}, 10240 \, \mathrm{m^{-1}}] $ and  $ f_{1} = 0$} \label{FIGdampl2f2}
\end{figure}

To investigate which range for the quadratic damping coefficient according to Eq. (\ref{damp1}) produces satisfactory wave damping, waves with wavelength $ \lambda = 4\, \mathrm{m} $ and height $ 0.16 \, \mathrm{m} $ are investigated. Only quadratic damping is considered, so $ f_{1} = 0 $. Simulations are performed for damping coefficient $ f_{2} $  between $ f_{2} \in [0.625 \, \mathrm{m^{-1}}, 10240 \, \mathrm{m^{-1}}] $. 
The recorded surface elevations near the end of the damping layer in Fig. \ref{FIGdampl2f2} show that the damping is strongly influenced by the choice of $ f_{2} $, while the wave phase and period remain nearly unchanged.

\begin{figure}[H]
\includegraphics[width=\factor\linewidth]{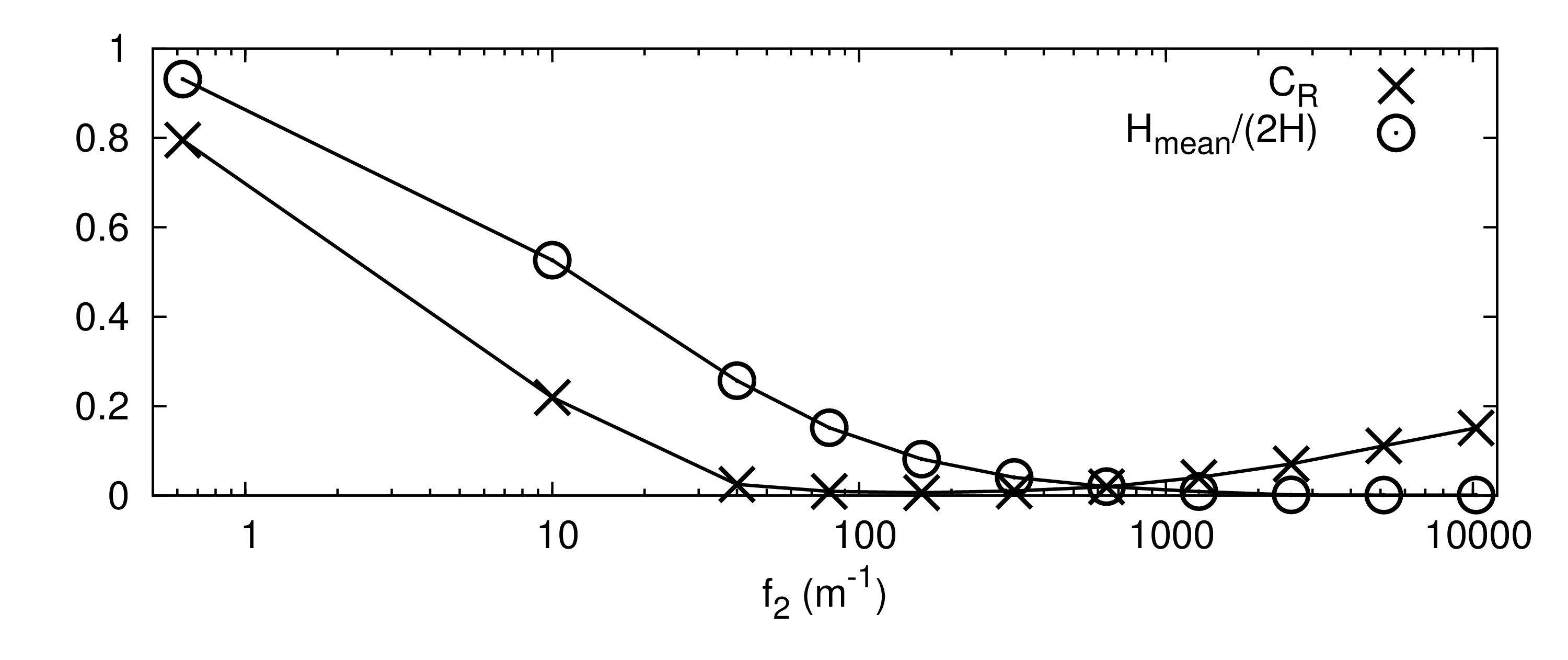}
\caption{Mean wave height $ H_{{\rm mean}} $ recorded at $x = 5.75\lambda$ scaled by twice the height $H$ of the undamped wave and reflection coefficient $ C_{\mathrm{R}} $  over damping coefficient $ f_{2} $, while $ f_{1} = 0 $} \label{FIGdampl2f2meanH}
\end{figure}

Figure  \ref{FIGdampl2f2meanH} shows how a variation of $ f_{2} $  affects the reflection coefficient $ C_{\mathrm{R}} $  and the mean wave height $H_{{\rm mean}} $ recorded in close vicinity to the boundary to which the damping layer is attached.  

The results show the same trends as the ones in Sect. \ref{l2f1}. For the given fade-in function and damping layer thickness, there is an optimum for $f_{2}$ so that the wave reflections propagating back into the solution domain are minimized. The best damping was achieved for  $f_{2} =160 \, \mathrm{m^{-1}}$, in which case the effects of wave reflections on the wave height within the solution domain are less than $ 0.7\% $.
The effects of wave reflections increase the further $f_{2}$ deviates from the optimum. For smaller $f_{2}$ values, the waves are reflected mainly at the domain boundary, for larger $f_{2}$ values the reflection occurs mainly at the entrance to the damping zone.
Significant reflections ($ C_{\mathrm{R}}>2\%$) in form of partial standing waves appear for roughly $ f_{2}< 80 \, \mathrm{m^{-1}} $ and $ f_{2}> 640 \, \mathrm{m^{-1}} $. 
Compared to the linear damping functions from the previous section, the use of quadratic damping functions does not offer a significant improvement in damping quality. With an optimal setup, both approaches provide roughly the same damping quality if the setup is optimized. However, the range of wave frequencies that are damped satisfactorily is narrower for quadratic damping.

\section{Influence of Computational Mesh on Achieved Damping}
\label{gridinf}
\begin{figure}[H]
\includegraphics[width=\factor\linewidth]{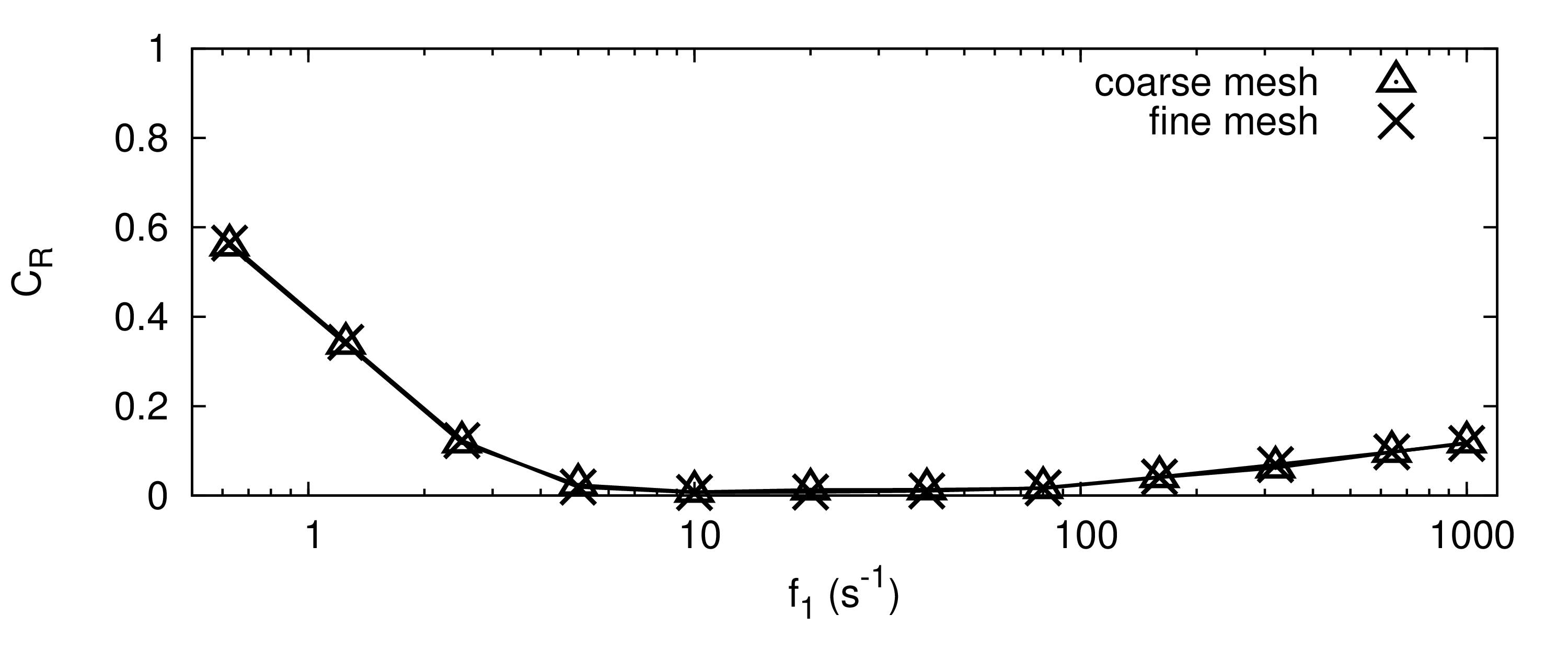}
\caption{Reflection coefficient $ C_{\mathrm{R}} $  over damping coefficient $ f_{1} $ for coarse and fine mesh simulations, with $ f_{2} = 0 $} \label{FIGdampl2f2_grid}
\end{figure}

The simulations  from Sect. \ref{l2f1} are rerun with same setup except for a grid coarsened by factor $2$. Thus whereas the fine mesh simulations discretize the wave with $100$ cells per wavelength and $16$ cells per wave height, the coarse mesh simulations have $50$ cells per wavelength and $8$ cells per wave height. All coarse grid reflection coefficients differ from their corresponding fine grid reflection coefficient by $ C_{\mathrm{R,\, coarse}} = C_{\mathrm{R,\, fine}} \pm 0.8\% $. This is also visible in Fig. \ref{FIGdampl2f2_grid}. Thus for sufficient resolution, the damping effectiveness can be considered grid-independent. This is expected since the damping-related terms in Eqs. (\ref{navier_stokes}) to (\ref{quad}) do not depend on cell volume. The used grids in this study are thus adequate.

\section{Influence of the Thickness of the Damping Layer}
\label{lengthxd}
\begin{figure}[H]
\includegraphics[width=\halffactor\linewidth]{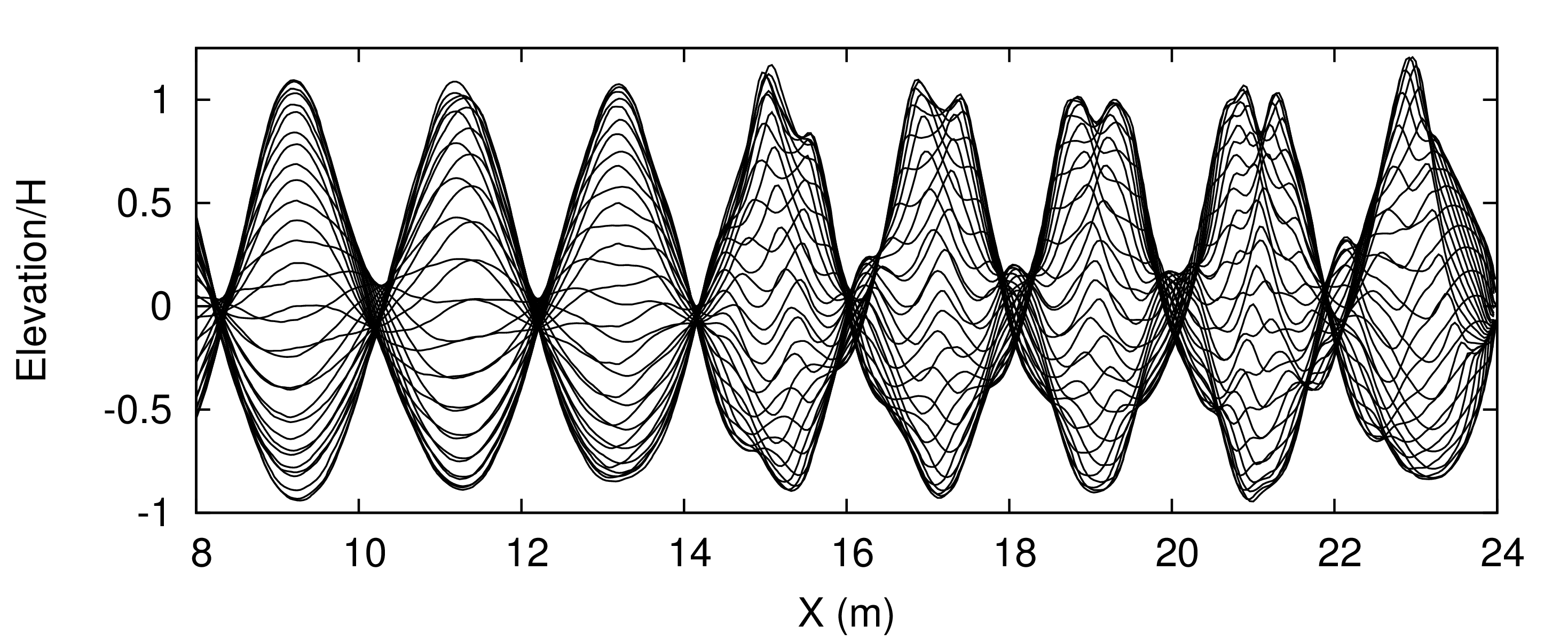}\\
\includegraphics[width=\halffactor\linewidth]{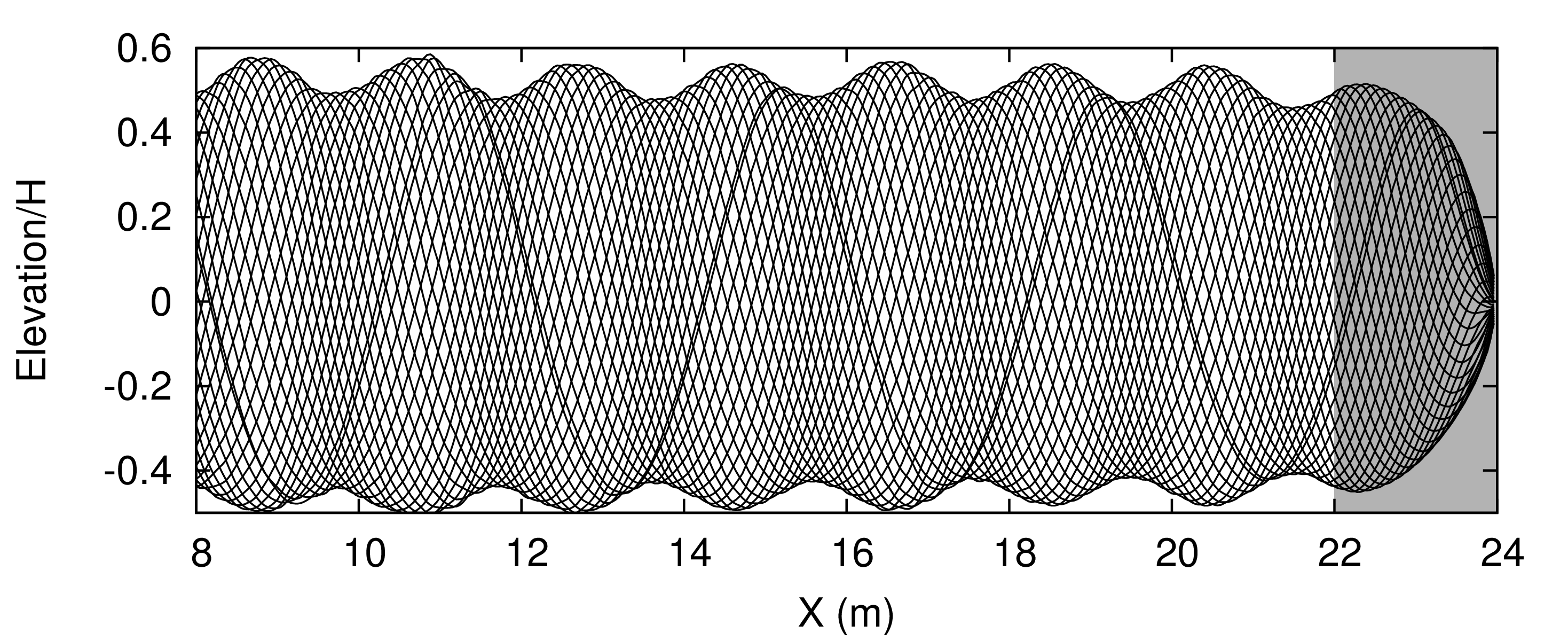}\\
\includegraphics[width=\halffactor\linewidth]{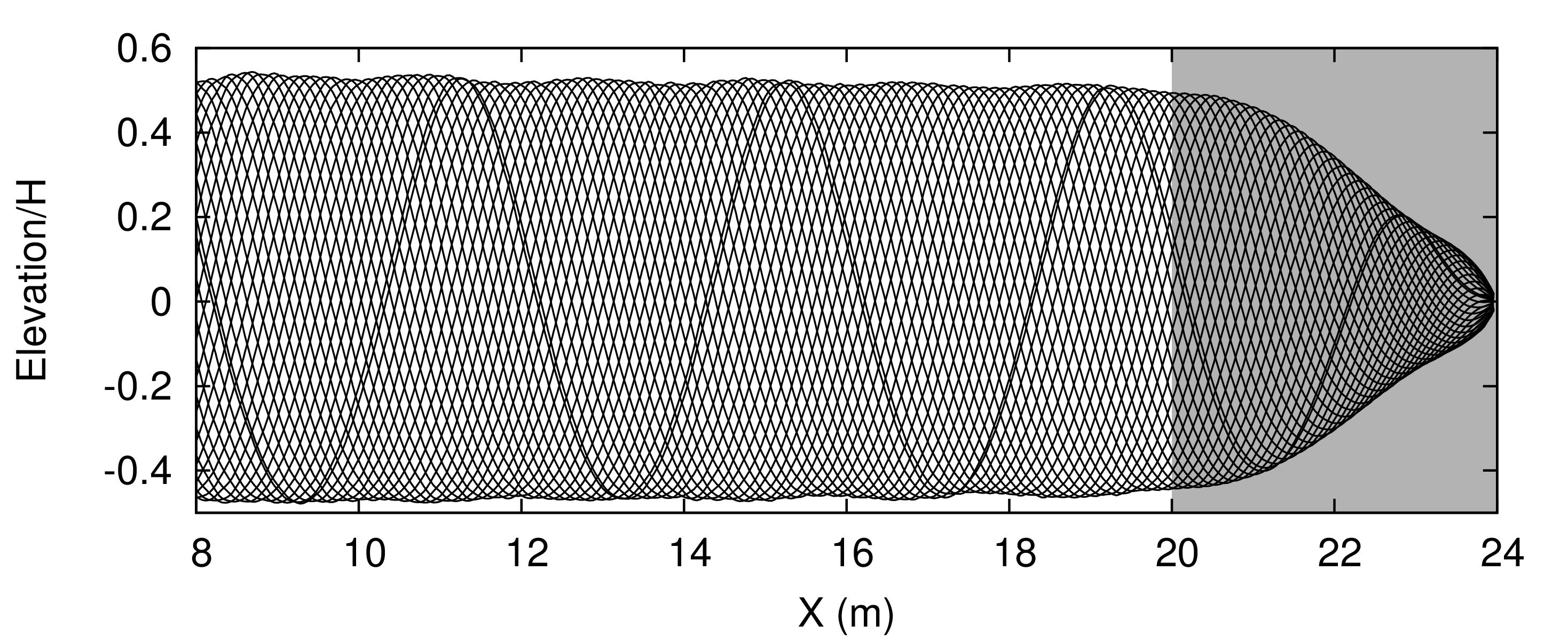}\\
\includegraphics[width=\halffactor\linewidth]{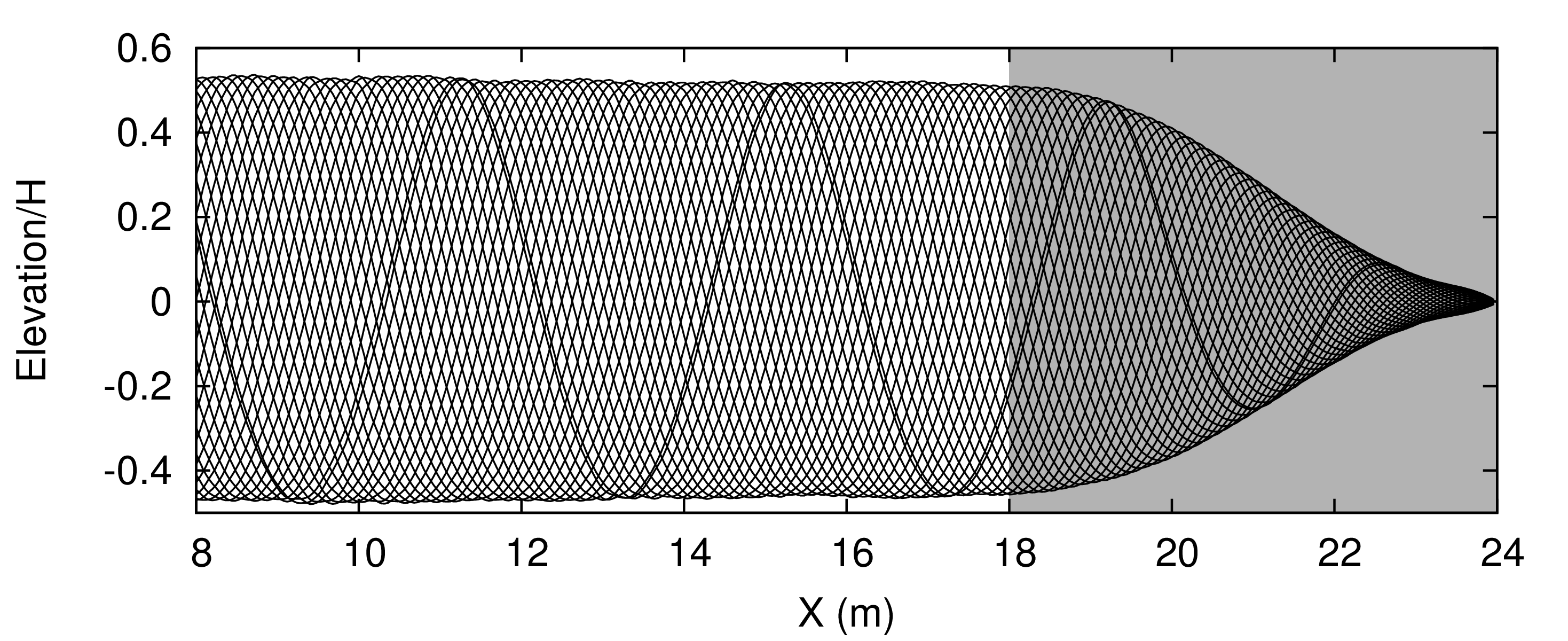}\\
\includegraphics[width=\halffactor\linewidth]{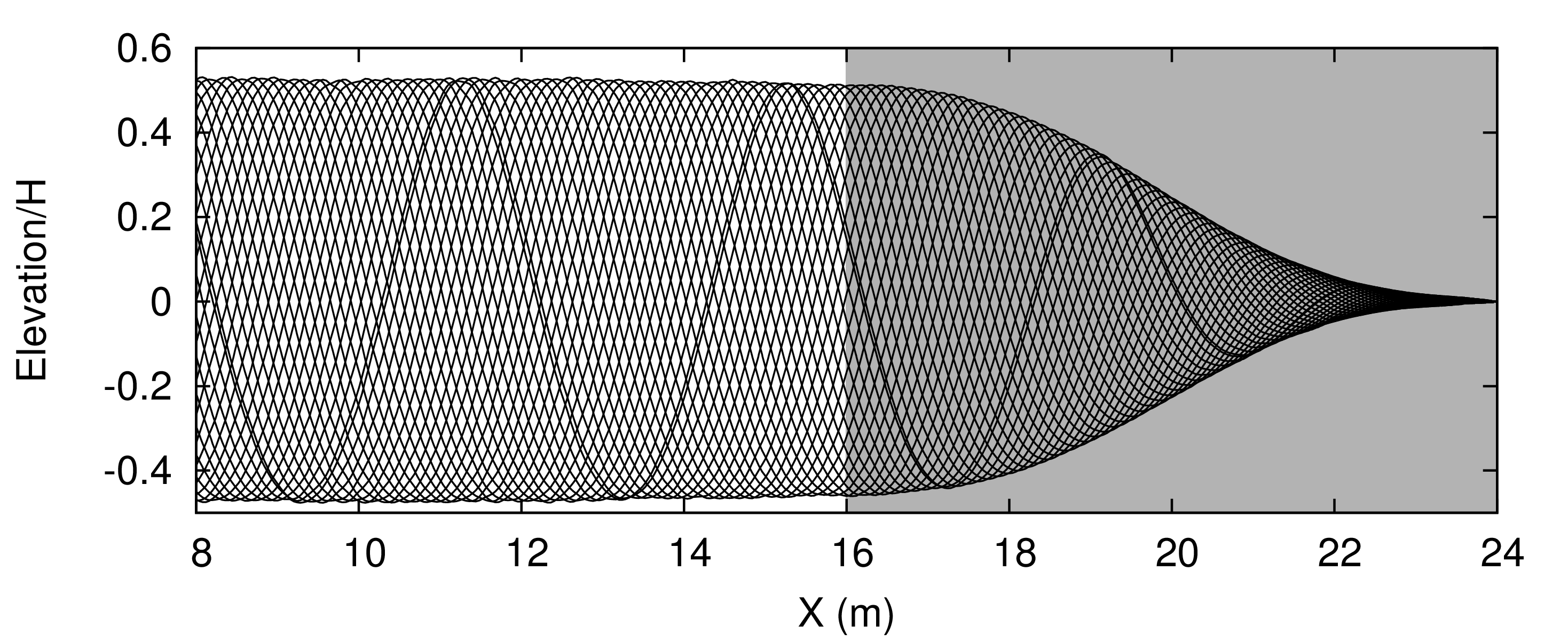}\\
\includegraphics[width=\halffactor\linewidth]{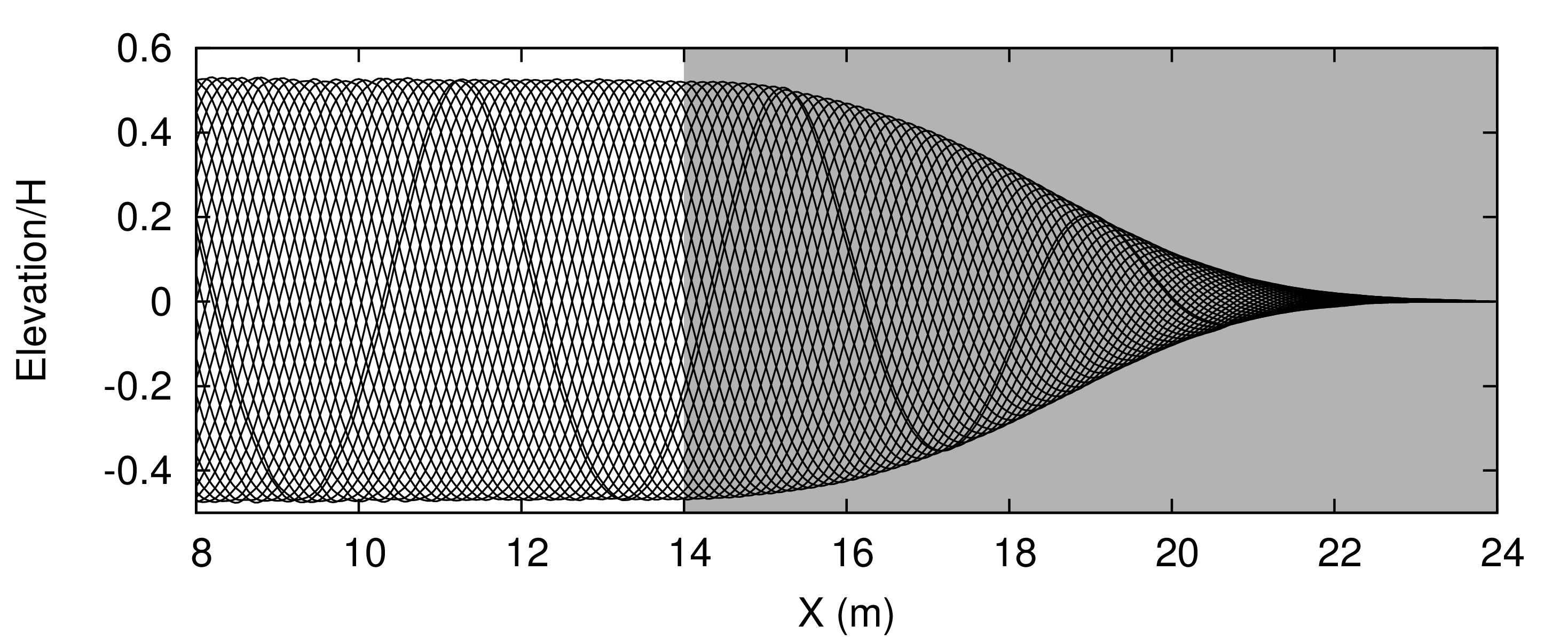}
\caption{Free surface elevation over $x$-location in tank shown for $40$ equally spaced time instances over one period starting at $t = 16\, \mathrm{s} $; from top to bottom  $ x_{{\rm d}} = 0\lambda, 0.5\lambda, 1.0\lambda, 1.5\lambda, 2.0\lambda, 2.5\lambda $; the damping zone is depicted as shaded gray} \label{FIGdamplf1surfelev}
\end{figure}

\begin{figure}[H]
\includegraphics[width=\factor\linewidth]{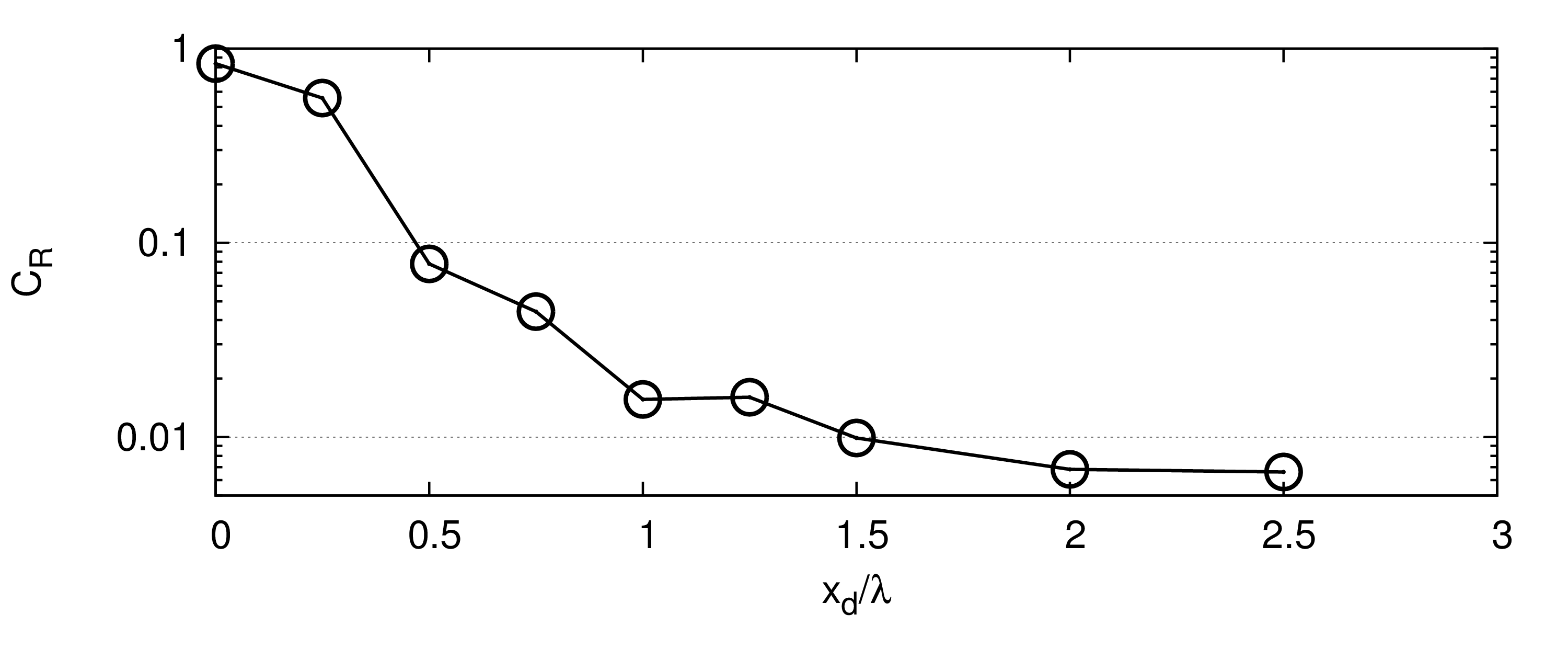}
\caption{Reflection coefficient $ C_{ \mathrm{R}} $ over damping zone thickness $x_{ \mathrm{d}}$ } \label{FIGdamplf1_CR}
\end{figure}

The simulation for  $\lambda = 4.0\, \mathrm{m}$, $H = 0.16\, \mathrm{m}$ and $f_{1} = 10 \, \mathrm{s^{-1}}$ from Sect. \ref{l2f1} is repeated for $x_{{\rm d}} =  0\lambda,  0.25\lambda, 0.5\lambda,  0.75\lambda, 1.0\lambda,  1.25\lambda, 1.5\lambda, 2.0\lambda, 2.5\lambda$ with otherwise the same setup.
The evolution of the free surface elevation in the tank over time in Fig. \ref{FIGdamplf1surfelev} shows that  $x_{{\rm d}}$ has a strong influence on the achieved damping.
Setting  $ x_{{\rm d}} = 0\lambda $ deactivates the damping and produces at first a nearly perfect standing wave, which then degenerates due to the influence of the pressure outlet boundary, since prescribing hydrostatic pressure establishes an oscillatory in-/outflow of water through this boundary which disturbs the standing wave.
For $ x_{{\rm d}} = 0.5\lambda $, a strong partial  standing wave occurs, and for $ x_{{\rm d}} = 1.0\lambda $ only  slight  reflections are still observable $ C_{\mathrm{R}} \approx 1.6\%$. 
For larger $x_{{\rm d}}$,  the influence of wave reflections continues to decrease. This is evident from the plot of $ C_{\mathrm{R}} $ for the simulations shown in Fig. \ref{FIGdamplf1_CR}. 

Subsequently, the simulations from Sect. \ref{l2f1} have been rerun with the damping thickness set to $ x_{\mathrm{d}} = 1\lambda$. Comparing the resulting curves for  $ C_{ \mathrm{R}} $ over $ f_{1} $ shows that increasing $ x_{\mathrm{d}} $ not only improves the damping quality; it also  widens the range of damping coefficients for which satisfactory damping is obtained; thus the wave damping then becomes less sensitive to $ \omega $ the more $ x_{\mathrm{d}} $ increases. However, this also increases the computational effort.
\begin{figure}[H]
\includegraphics[width=\factor\linewidth]{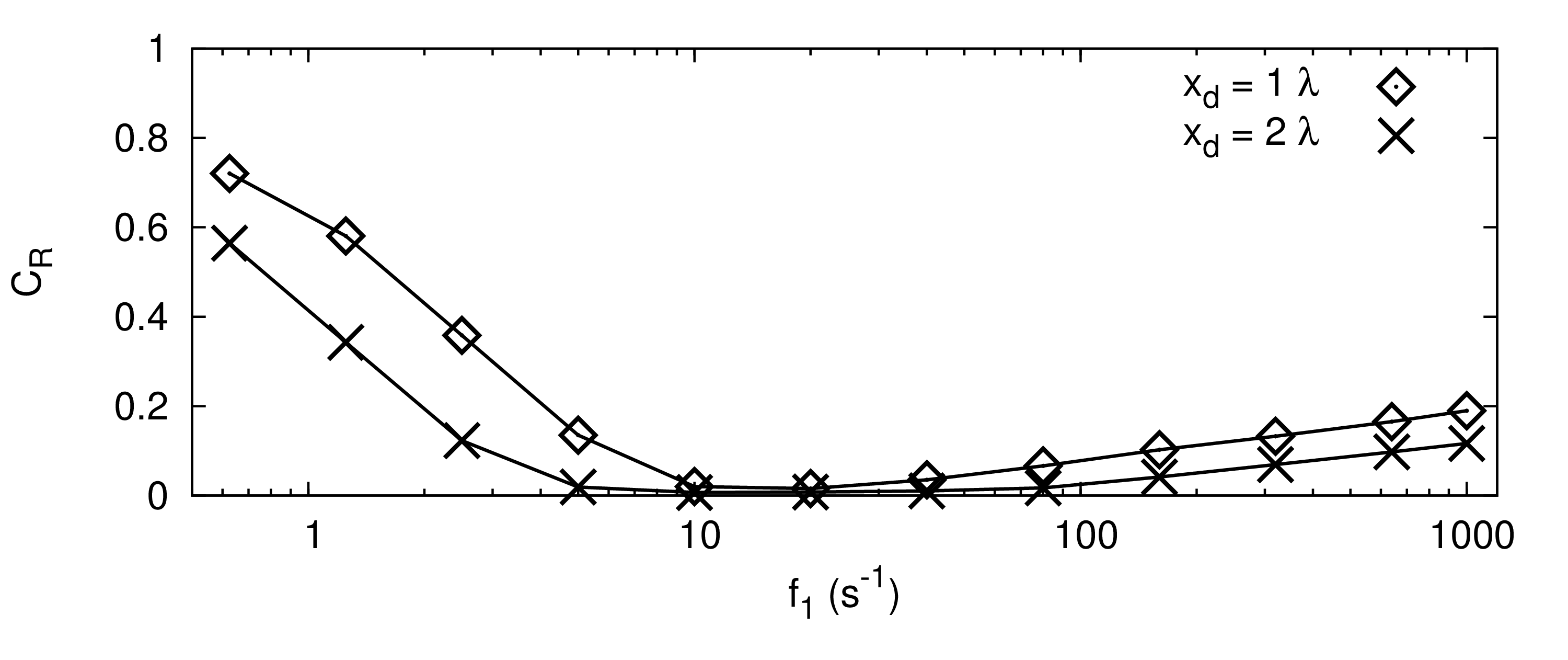}
\caption{Reflection coefficient $ C_{\mathrm{R}} $  over damping coefficient $ f_{1} $ for $ x_{\mathrm{d}} = 1\lambda$ and $ x_{\mathrm{d}} = 2\lambda$, while $ f_{2} = 0 $} \label{FIGdampl2f2_damplayer}
\end{figure}

The results show that, if the damping coefficients are set up close to the optimum and it is desired that  $ C_{\mathrm{R}} < 2\%$, then  $ x_{\mathrm{d}} = 1\lambda$ suffices. 
This knowledge is useful, since by reducing $ x_{\mathrm{d}}$ the computational domain can be kept smaller and thus the computational effort can be reduced. However, if better damping is desired or when complex flow phenomena are considered,  especially when irregular waves or wave reflections from bodies are present, then the damping layer thickness should be increased  to damp all wave components successfully. 
The present study suggests that a damping layer thickness of $1.5\lambda \leq x_{{\rm d}} \leq 2 \lambda$ can be recommended.

\section{Influence of Wave Steepness on Achieved Damping}
\label{heightinfl}
The simulations in this section are based on those from  Sect. \ref{l2f1}, i.e.  $\lambda = 4.0\, \mathrm{m}$, $H = 0.16\, \mathrm{m}$  and varying $f_{1} $ in the range $[0.625 \, \mathrm{s^{-1}}, 1000 \, \mathrm{s^{-1}}]$. The simulations were rerun with the same setup except for two modifications: The wave height was changed to $H = 0.4\, \mathrm{m}$, resulting in a steepness of  $H/\lambda = 0.1$ instead of the previous $H/\lambda = 0.04$. Furthermore, the grid was adjusted to maintain the same number of cells per wave height  as well as per wavelength, so that both results are comparable.

\begin{figure}[H]
\includegraphics[width=\factor\linewidth]{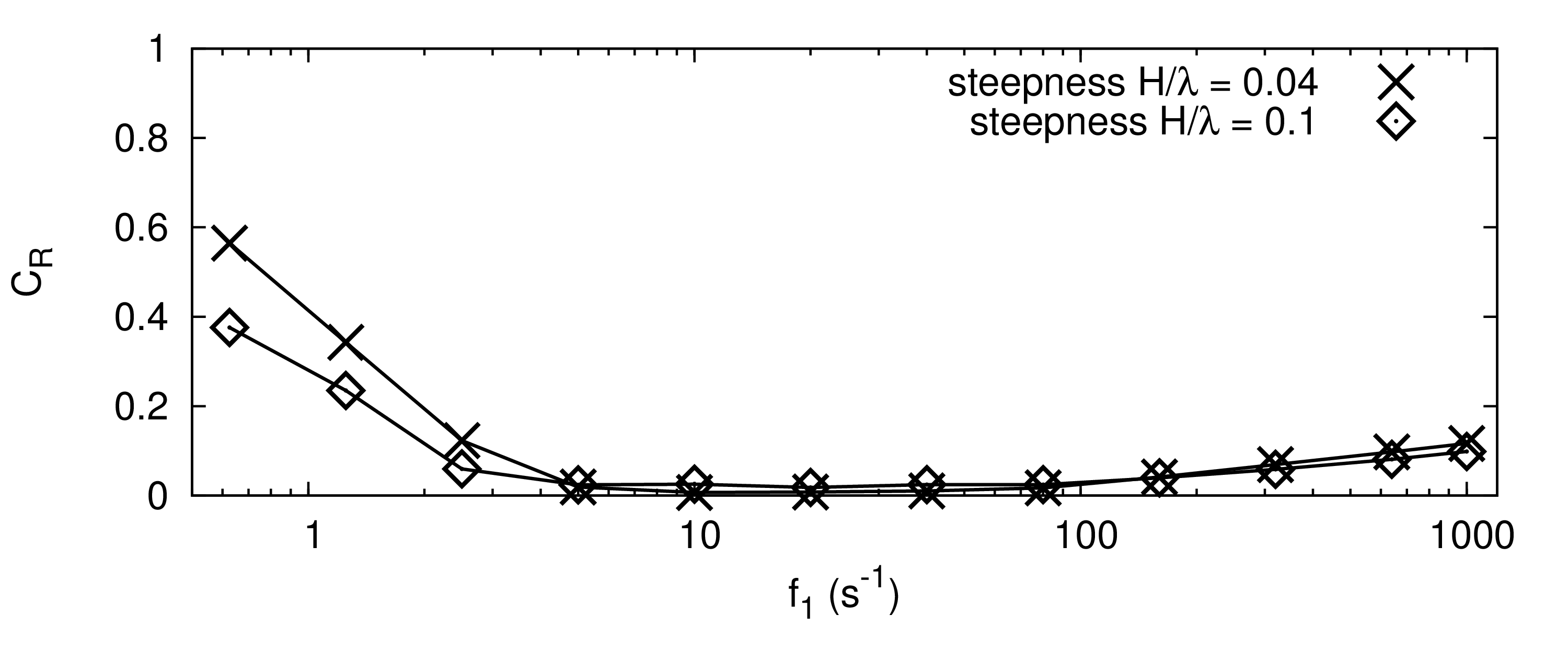}
\caption{Reflection coefficient $ C_{\mathrm{R}} $  over damping coefficient $ f_{1} $ for waves of same period but differing steepness $ H/\lambda = 0.04 $ and  $ H/\lambda = 0.1 $, while $ f_{2} = 0 $} \label{FIGsteepnessinfl}
\end{figure}

As can be seen from Figure \ref{FIGsteepnessinfl}, the influence of the increased wave steepness is comparatively small, except for the cases with significantly smaller than optimum damping ($ f_{1} \leq 2.5  \, \mathrm{s^{-1}}$). 
For the rest of the range, i.e. $5 \leq f_{1} \leq 1000 \, \mathrm{s^{-1}}$, the difference in reflection coefficients is only $ C_{\mathrm{R}}\left( H/\lambda = 0.1\right) =  C_{\mathrm{R}}\left( H/\lambda = 0.04\right) \pm 1.7\% $. Therefore, although the damping performs slightly better for waves of smaller steepness, the influence of wave steepness can be assumed negligible for most practical cases. 
If stronger wave steepness is considered and less uncertainty is required, then the thickness of the damping layer can be further increased to decrease $C_{\mathrm{R}}$ .

\section{The Scaling Law for Linear Damping}
\label{Scalel2f1}
In order to verify the assumed scaling law for linear damping from Sect. \ref{Lf1}, the simulation setup  with the best damping performance ($ f_{1} = 10.0  \, \mathrm{s^{-1}}$) from Sect. \ref{l2f1} was scaled geometrically and kinematically so that the generated waves are completely similar, for wavelengths $ \lambda = 0.04\, \mathrm{m}, 4\, \mathrm{m}, 400\, \mathrm{m} $ and thus corresponding heights of $H = 0.0016\, \mathrm{m}, 0.16\, \mathrm{m}, 16\, \mathrm{m}$. This corresponds to a realistic scaling, since geometrically scaling by $1:100$ is common in both experimental and computational model- and full scale investigations. 
As necessary requirement to obtain similar damping (i.e. similarity of $C_{\mathrm{R}}$ and surface elevation), the damping length $ x_{{\rm d}} $ is scaled directly proportional to the wavelength, according to Eq. (\ref{xdscale}). 

\begin{figure}[H]
\includegraphics[width=\factor\linewidth]{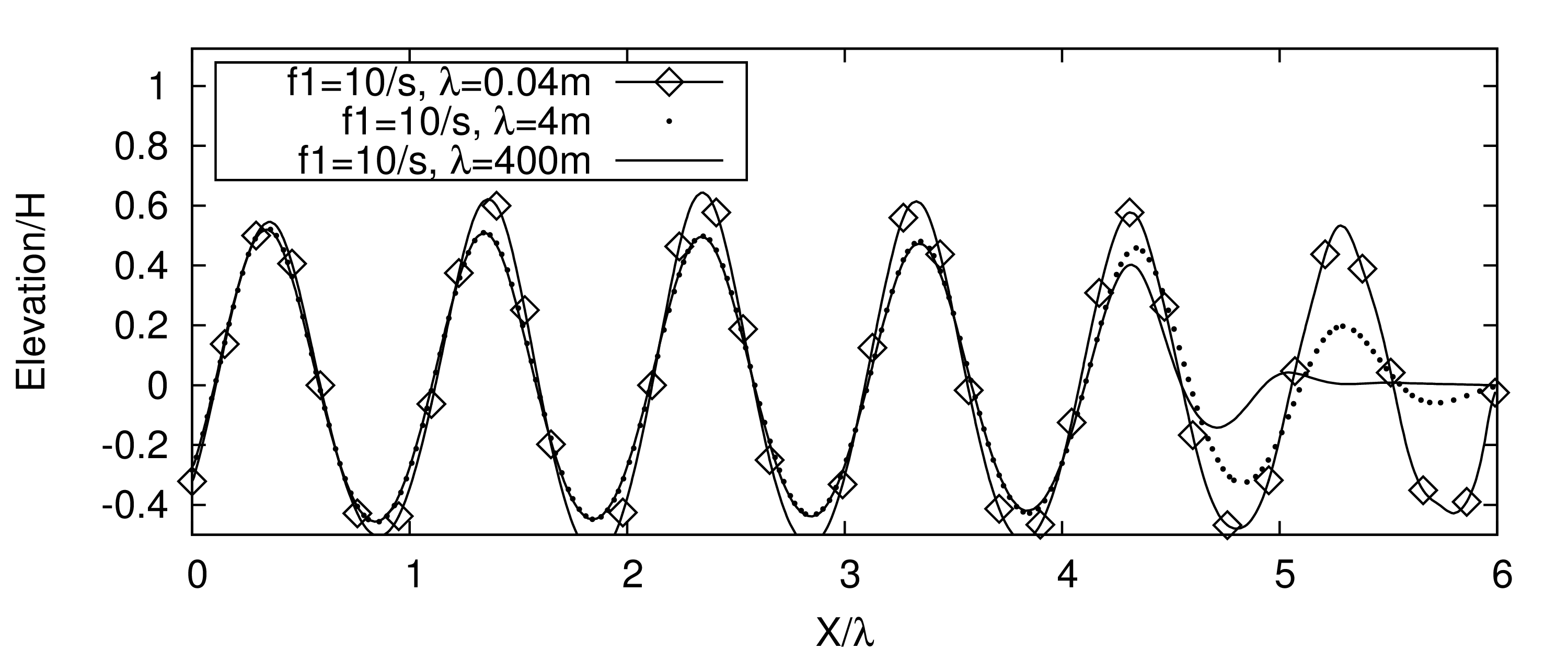}
\caption{Surface elevation in the whole domain after $ \approx 12.6 $ periods; for similar waves with wavelengths $\lambda = 0.04\, \mathrm{m}, 4\, \mathrm{m}, 400\, \mathrm{m} $;\textit{ no scaling  of  $f_{1}$}, thus $ f_{1} = 10  \, \mathrm{s^{-1}}$ and  $ f_{2} = 0$ for all cases}
 \label{FIGdampScalel2f1freesurf_unscaled}
\end{figure}

\begin{figure}[H]
\includegraphics[width=\factor\linewidth]{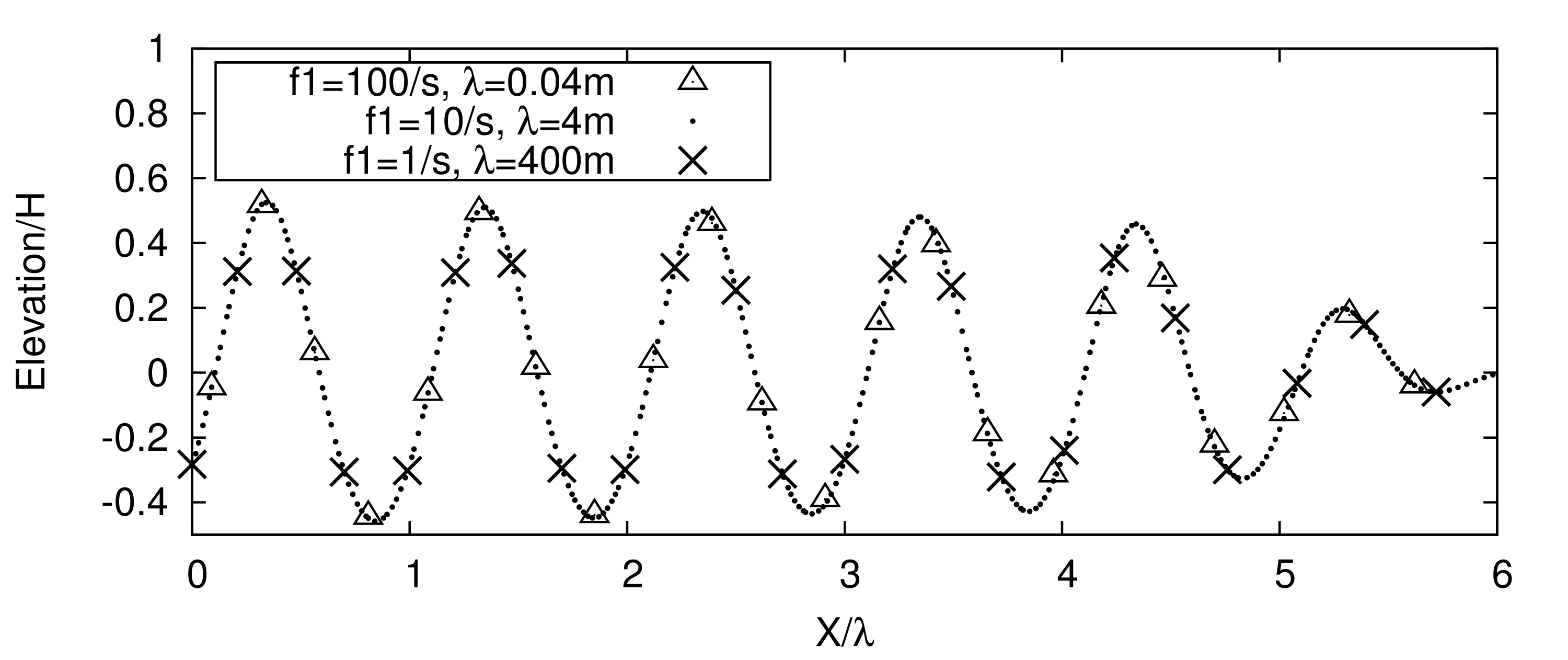}
\caption{Surface elevation in the whole domain after $ \approx 12.6 $ periods; for similar waves with wavelengths $\lambda = 0.04\, \mathrm{m}, 4\, \mathrm{m}, 400\, \mathrm{m} $; \textit{ scaling of  $f_{1}$ according to Eq.  (\ref{omegascale})}, thus  $ f_{1} = 10 \, \mathrm{s^{-1}}, 3.16 \, \mathrm{s^{-1}}, 1  \, \mathrm{s^{-1}}$ and  $ f_{2} = 0$}
 \label{FIGdampScalel2f1freesurf_scaled}
\end{figure}

At first, the simulations are run with no scaling of $f_{1}$, so that  $f_{1} = 10 \, \mathrm{s^{-1}}$ in all cases. Figure \ref{FIGdampScalel2f1freesurf_unscaled} shows the free surface in the tank after the simulations. In the vicinity of the inlet ($0 < x/\lambda < 1 $) the effects of wave reflections have not yet fully established, thus the differences between the curves are small. This shows that the wave-maker generated similar waves in the three simulations. 
As the plot shows an arbitrarily selected time instant, the effects of wave reflections  cannot be judged from it, since the partial standing waves were not captured at their maximum amplification. Therefore, for  $  \lambda = 400\, \mathrm{m}$ no noticeable difference to the reference case $  \lambda = 4\, \mathrm{m}$ is seen outside the damping zone ($0 < x/\lambda < 4 $), whereas obvious reflections are visible for $  \lambda = 0.04\, \mathrm{m}$. However regarding the surface elevation within the damping zone ($ 4 < x/\lambda < 6 $), the three curves differ considerably, and thus the wave damping also does not act in a similar fashion.  A close look at the surface elevation in the damping zones shows that wave reflections will mostly occur at domain boundary $ x/\lambda = 6 $ for $ \lambda=0.04\, \mathrm{m} $, in contrast to $ \lambda=400\, \mathrm{m} $, where they will mostly occur close to the entrance to the damping zone at $ x/\lambda = 4 $.
Subsequently, based on the findings in Sect. \ref{Lf1}, the simulations are rerun with $f_{1}$ scaled with wave frequency $ \omega $ according to Eq. (\ref{omegascale}), based on  reference case $ \lambda_{\mathrm{ref}} = 4\, \mathrm{m}$.  It is apparent from Fig. \ref{FIGdampScalel2f1freesurf_scaled}, that the surface elevations are similar everywhere in the tank and that the proposed scaling law is correct.
Table \ref{TABf1scal} compares the reflection coefficients  $C_{\mathrm{R}}$ for both unscaled and correctly scaled $ f_{1}$. 
When $f_{1}$ is held constant, an up- or down-scaling of the wave with factor $100$ will lead to a significant increase in reflections. Therefore without adjusting  $f_{1}$, it is not possible to obtain similar damping in model- and full scale.

If instead $f_{1}$ is scaled according to Eqs. (\ref{xdscale}) and (\ref{omegascale}),  $C_{\mathrm{R}}$ stays nearly the same; slight fluctuations of  $C_{\mathrm{R}}$ are due to the scheme for obtaining $C_{\mathrm{R}}$. Therefore, the damping quality can be reliably reproduced when the scaling law is applied. 

This is even evident in the $C_{\mathrm{R}}$ values for fixed $f_{1}$: For the scaled waves considered, $ f_{1} $ is either roughly $10$ times larger ($ \lambda = 400\, \mathrm{m} $) or smaller ($ \lambda = 0.04\, \mathrm{m} $) than the optimum value; compared with the   $10$ times larger or smaller than optimal $ f_{1} $ values from Fig. \ref{FIGdampl2f1meanH}, the $C_{\mathrm{R}}$ values coincide.
\begin{table}[H]
\caption{Reflection coefficient $C_{\mathrm{R}}$ for unscaled and correctly scaled simulations}\label{TABf1scal}
\begin{center}
\tabcolsep=0.25cm
\begin{tabular}{c| c c}
\hline
$ \lambda $ & $C_{\mathrm{R}} (\mathrm{fixed}\ f_{1})$ &  $C_{\mathrm{R}} (\mathrm{scaled}\ f_{1} )$\\ [0.5ex]
\hline
$ 0.04\, \mathrm{m} $ & $39.8\%$ &  $1.1\%$ \\ 
\hline 
$ 4\, \mathrm{m} $ & $0.7\%$ & $0.7\%$  \\ 
\hline 
$ 400\, \mathrm{m} $ &  $2.4\%$ & $0.6\%$ \\ 
\hline 
\end{tabular} 
\end{center}
\end{table}
\vspace*{0.2cm}
\fbox {
    \parbox{\linewidth}{
\textbf{Practical Recommendation}\\
When using linear damping according to Eqs. (\ref{damp1})  and (\ref{damp2}), the following damping setup can be recommended based on the results from Sects. \ref{l2f1} to \ref{irrwav}:
\begin{equation}
f_{1} = \Psi_{1} \omega  \quad ,
\label{f1_law}
\end{equation}
with $ \Psi_{1} = \pi$, wave frequency $\omega$, $f_{2} = 0$,  $x_{{\rm d}}=2\lambda $ and $n=2$.
    }
}

\section{The Scaling Law for Quadratic Damping}
\label{Scalel2f2}
In order to verify the assumed scaling law for quadratic damping from Sect. \ref{Lf1}, the simulation setup  with the best damping performance ($ f_{2} = 160.0 \, \mathrm{m^{-1}}$) from Sect. \ref{l2f2} was scaled geometrically and kinematically so that the generated waves are completely similar, for wavelengths $ \lambda = 0.04\, \mathrm{m}, 4\, \mathrm{m}, 400\, \mathrm{m} $ and thus corresponding heights of $H = 0.0016\, \mathrm{m}, 0.16\, \mathrm{m}, 16\, \mathrm{m}$. As in the previous section, the damping length $ x_{{\rm d}} $ is scaled directly proportional to the wavelength, according to Eq. (\ref{xdscale}). 

\begin{figure}[H]
\includegraphics[width=\factor\linewidth]{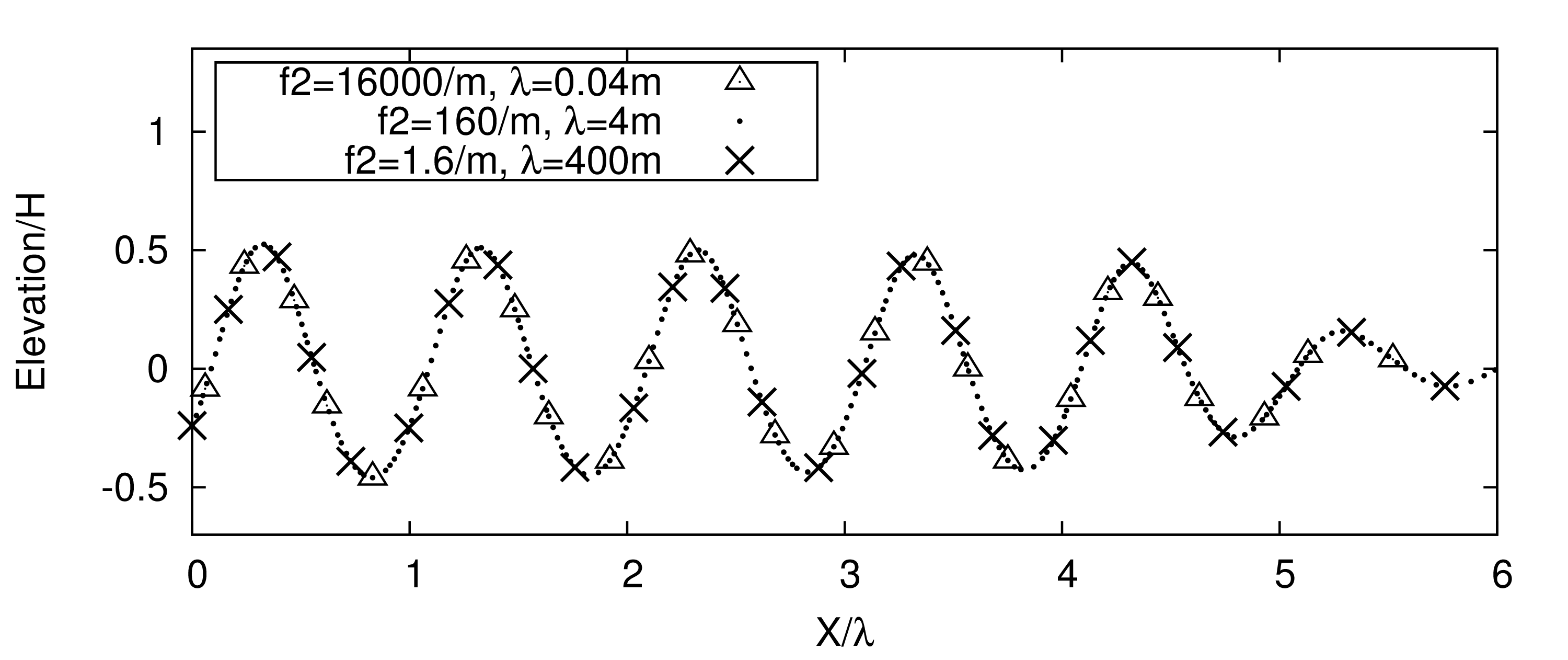}
\caption{Surface elevation in the whole domain after $ \approx 12.6 $ periods; for similar waves with wavelengths $\lambda =0.04\, \mathrm{m}, 4\, \mathrm{m}, 400\, \mathrm{m} $; \textit{ scaling of  $f_{2}$ according to Eq. (\ref{omegascale2})}, thus  $ f_{2} = 160 \, \mathrm{m^{-1}}, 16 \, \mathrm{m^{-1}}, 1.6 \, \mathrm{m^{-1}} $ and  $ f_{1} = 0$}
 \label{FIGdampScalel2f2freesurf2}
\end{figure}

At first the simulations are run with no scaling of $f_{2}$, so that  $f_{2} = 160 \, \mathrm{m^{-1}}$ in all cases. 
Subsequently, based on the findings in Sect. \ref{Lf1}, the simulations are rerun with $f_{2}$ scaled with wave frequency $ \omega $ squared according to Eq. (\ref{omegascale2}), based on  reference case $ \lambda = 4\, \mathrm{m}$.  It is apparent from Fig. \ref{FIGdampScalel2f2freesurf2}, that the surface elevations are similar everywhere in the tank and that the proposed scaling law is correct.
Table \ref{TABf2scal} compares the reflection coefficients  $C_{\mathrm{R}}$ for both unscaled and correctly scaled $ f_{2}$. 
When $f_{2}$ is held constant, an up- or down-scaling of the wave with factor $100$ will lead to a significant increase in reflections.
If instead $f_{2}$ is scaled according to Eqs. (\ref{xdscale}) and (\ref{omegascale}),  $C_{\mathrm{R}}$ stays nearly the same; slight fluctuations of  $C_{\mathrm{R}}$ are due to the scheme for obtaining $C_{\mathrm{R}}$. This shows that the damping quality can be reliably reproduced when the scaling law is applied. 

\begin{table}[H]
\caption{Reflection coefficient $C_{\mathrm{R}}$ for unscaled and correctly scaled simulations}\label{TABf2scal}
\begin{center}
\tabcolsep=0.25cm
\begin{tabular}{c| c c}
\hline
$ \lambda $ & $C_{\mathrm{R}} (\mathrm{fixed}\ f_{2})$ &  $C_{\mathrm{R}} (\mathrm{scaled}\ f_{2} )$\\ [0.5ex]
\hline
$ 0.04\, \mathrm{m} $ & $65.9\%$ &  $1.1\%$ \\ 
\hline 
$ 4\, \mathrm{m} $ & $0.6\%$ & $0.6\%$  \\ 
\hline 
$ 400\, \mathrm{m} $ &  $17.7\%$ & $0.5\%$ \\ 
\hline 
\end{tabular} 
\end{center}
\end{table}

The present results show that if the damping parameters are not adjusted when performing model- and full-scale simulations, then unsatisfactory damping can be expected. 

Compared to linear wave damping, quadratic damping has a narrower range of wave frequencies for which satisfactory damping is obtained. 
Therefore, since the damping performance is more sensitive to changes in wave frequency for quadratic (or higher order) damping functions, and since with optimum set up the damping performance is the same as with linear damping,  it is recommended to use linear damping functions. For this reasons, only linear damping was examined in the following Sects. \ref{irrwav} and \ref{kcs}.\vspace*{0.2cm}\\
\fbox {
    \parbox{\linewidth}{
\textbf{Practical Recommendation}\\
When using quadratic damping according to Eqs. (\ref{damp1})  and (\ref{damp2}), the following damping setup can be recommended based on the results from Sect. \ref{l2f2}:
\begin{equation}
f_{2} = \Psi_{2}  \lambda^{-1}  \quad ,
\label{f2_law}
\end{equation}
with  $ \Psi_{2} = 2\pi \cdot 10^{2}$, wavelength $\lambda$, $f_{1} = 0$,  $x_{{\rm d}}=2\lambda $ and $n=2$.
    }
}

\section{Damping of Irregular Free-Surface Waves}
\label{irrwav}
The previous results indicate that, to simulate irregular waves,  the damping setup should be based on  the wave component with the longest wavelength. If the damping setup is optimal for the longest wave component, then on the one hand the damping coefficient for all shorter wave components will be smaller then optimal, which has a negative effect on the damping performance for the corresponding wave component as seen in Sect. \ref{l2f1}. On the other hand, the damping layer thickness relative to the wavelength will be larger for these wave components, which has a positive effect on the damping performance as shown in Sect. \ref{lengthxd}. From the previous results, the latter effect is expected be the prevailing influence on the damping quality. Furthermore, the shorter wave components are discretized with less cells per wavelength, so the discretization errors are stronger for these wave components, which provides a faster wave energy dissipation for these components which is also beneficial in this respect. 

Exemplarily,  linear damping of three superposed wave components with same steepness $ \frac{H_{i}}{\lambda_{i}} = 0.03 $ and wavelengths differing by $400\%$ is demonstrated in the following. At the wave-maker, the following surface elevation is prescribed
\begin{equation}
\eta(t) = \sum_{i=1}^{3} \frac{H_{i}}{2} \cos(-\omega_{i} (t + 1.19\, \mathrm{s}) + \phi_{i} )\quad ,
\label{irrEQ}
\end{equation}
with wave height $H_{i}  $, wave frequencies $ \omega_{i} $, phase shifts $  \phi_{i} $ and time $ t $. The actual parameters are shown in Table \ref{TABirr}.
\begin{table}[H]
\caption{Parameters of the wave components}\label{TABirr}
\begin{center}
\tabcolsep=0.25cm
\begin{tabular}{c| c c c c }
\hline
$i$ & $\lambda_{i} (\mathrm{m})$ & $H_{i} (\mathrm{m})$ & $\omega_{i} (\frac{ \mathrm{rad}}{\mathrm{s}})$ & $\phi_{i} (\mathrm{rad})$\\ 
\hline
$1$ & $4.0$ & $0.12$ & $3.926$ & $0.0$  \\ 
\hline 
$2$ & $2.0$ & $0.06$ & $5.551$ & $-0.9176$ \\ 
\hline 
$3$ & $1.0$ & $0.03$ & $7.851$ & $3.543$  \\ 
\hline 
\end{tabular} 
\end{center}
\end{table}
Velocity and volume fraction corresponding to Eq. (\ref{irrEQ}) are taken from linear wave theory and are applied at the inlet boundary $ x = 0 $ as linear superposition. The domain length is $L_{x} = 3\lambda_{1} $.
The grid dimensions in free surface zone are $ \Delta x = 0.0195\, \mathrm{m}$ and $ \Delta z = 0.0024 \, \mathrm{m}$ with a time step of $ \Delta t = 0.001\, \mathrm{s} $. 

\begin{figure}[H]
\includegraphics[width=\factor\linewidth]{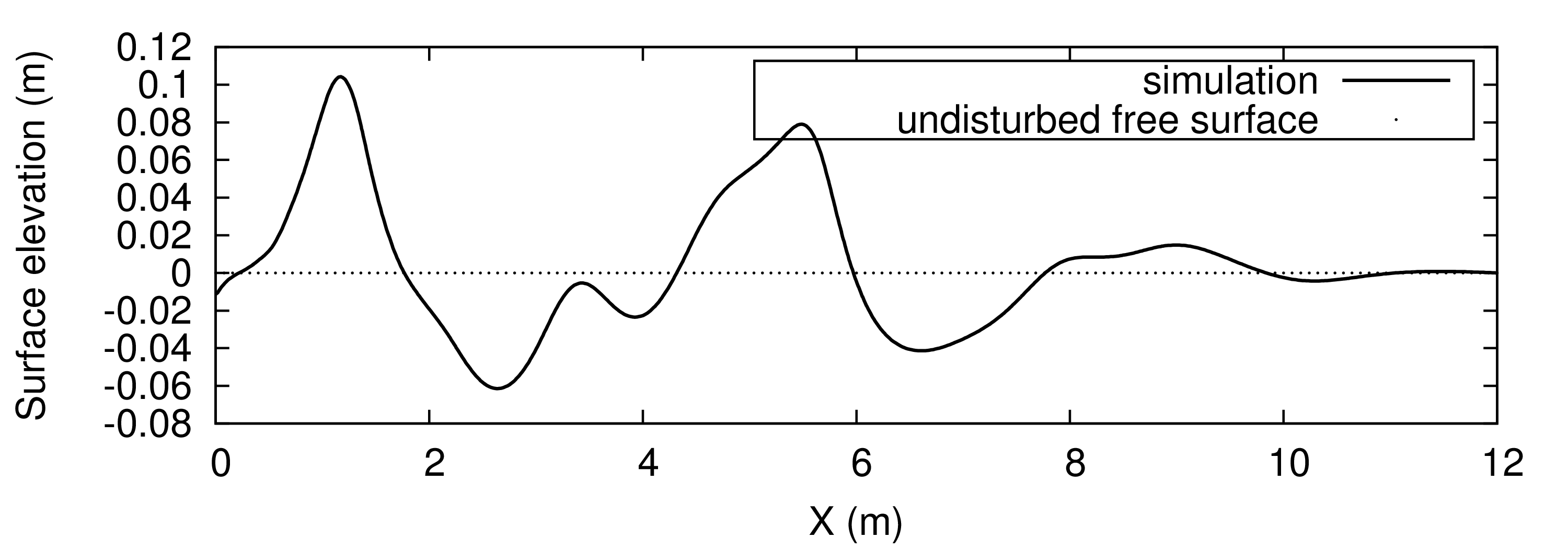}\\
\includegraphics[width=\factor\linewidth]{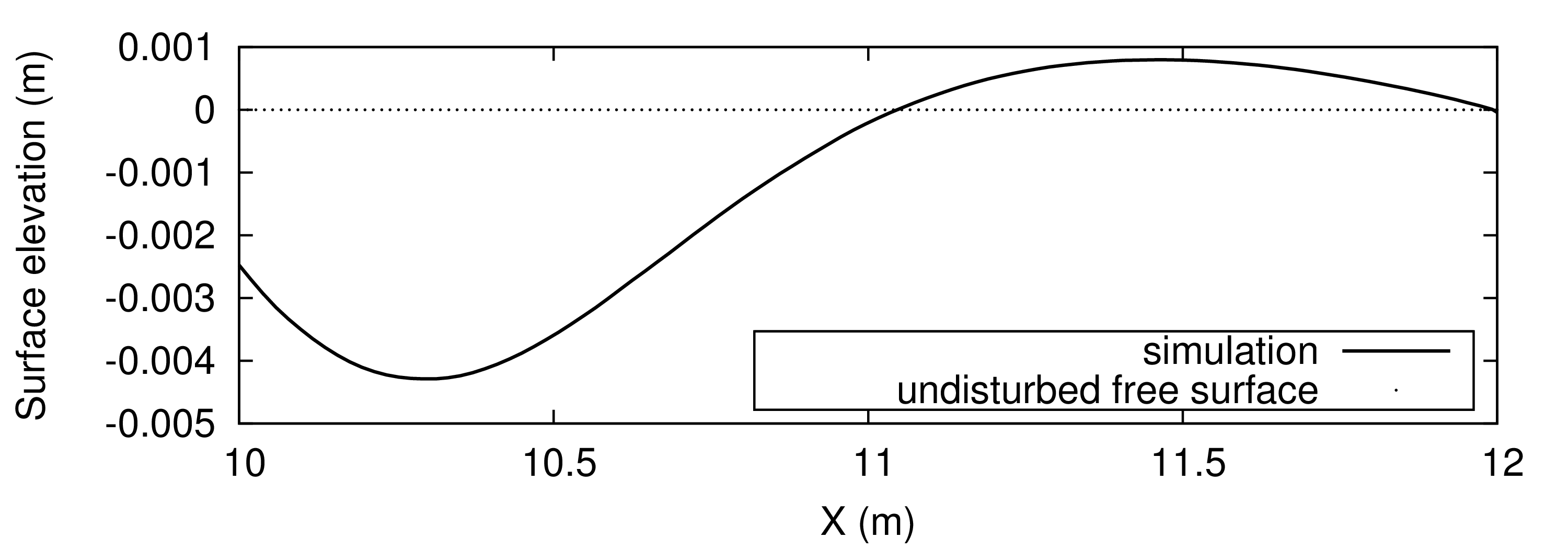}
\caption{Free surface elevation in the solution domain after $20\, \mathrm{s}$; top: whole domain; bottom: close up of the last part of the damping layer} \label{FIGirrSurf}
\end{figure}

The damping setup is based on the wave component with the longest wavelength according to Eq. (\ref{f1_law}): The damping layer thickness is $x_{d} = 2\lambda_{1}$ and the damping coefficients are $f_{1} = 20 \, \mathrm{s^{-1}}$ and   $f_{2} = 0$. With this setup, all wave components were damped satisfactorily without noticeable reflections. As can be seen from  Fig. \ref{FIGirrSurf}, the waves enter the damping layer  at $ x = 4 \, \mathrm{m}$ and  after propagating a distance of  $ 2\lambda_{1} $ into the damping layer, the wave height is reduced by two orders of magnitude. Thus linear damping seems well suited for the damping of irregular waves.  Although the approach from Sect. \ref{Lf1} to determine $C_{\rm R}$ cannot be applied in this case, the results from Sect. \ref{l2f1} show that for all wave components reflections will mostly occur at the domain boundary, and not at the entrance to the damping layer. Thus in this case, measuring the surface elevation in close vicinity to the corresponding domain boundary is enough to evaluate the damping performance.

However, for realistic wave spectra the above approach is not feasible, as these consist of wave components over a broad range of wavelengths. 
Thus selecting the component with the longest wavelength for the damping zone setup would not be practical, since this would require a very large damping zone and increase the total amount of grid cells and computational effort substantially (possibly one or more orders of magnitude if the mesh size is not changed).

For practical purposes, an efficient approach is outlined in the following. According to Lloyd (1989), most realistic sea spectra are narrow banded. This means that most wave energy is concentrated in a narrow band of wave frequencies around a peak frequency. Although larger wavelengths occur as well, these carry  a comparatively small amount of energy. Thus we propose to set up the wave damping based on the peak wave frequency. As seen in Sect. \ref{lengthxd} the range of frequencies that are damped can be modified with the damping zone thickness, so the broader-banded the spectrum is, the larger has $ x_{\rm d} $ to be.

Exemplarily, this approach is investigated for the widely used JONSWAP spectrum, which was originally developed by Hasselmann et al. (1973). Irregular waves are prescribed at the inlet boundary, with parameters peak wave period $T_{\mathrm{p}} = 1.6 \, \mathrm{s}$, significant wave height $H_{\mathrm{s}} = 0.12\, \mathrm{m}$, peak shape parameter $ \gamma = 3.3 $, spectral width parameter $ \sigma = 0.07 $ for $ \omega \leq \omega_{\mathrm{p}}$ and $ \sigma = 0.09 $ for $ \omega > \omega_{\mathrm{p}}$. The wave spectrum was discretized into $100$ components.

\begin{figure}[H]
\includegraphics[width=\factor\linewidth]{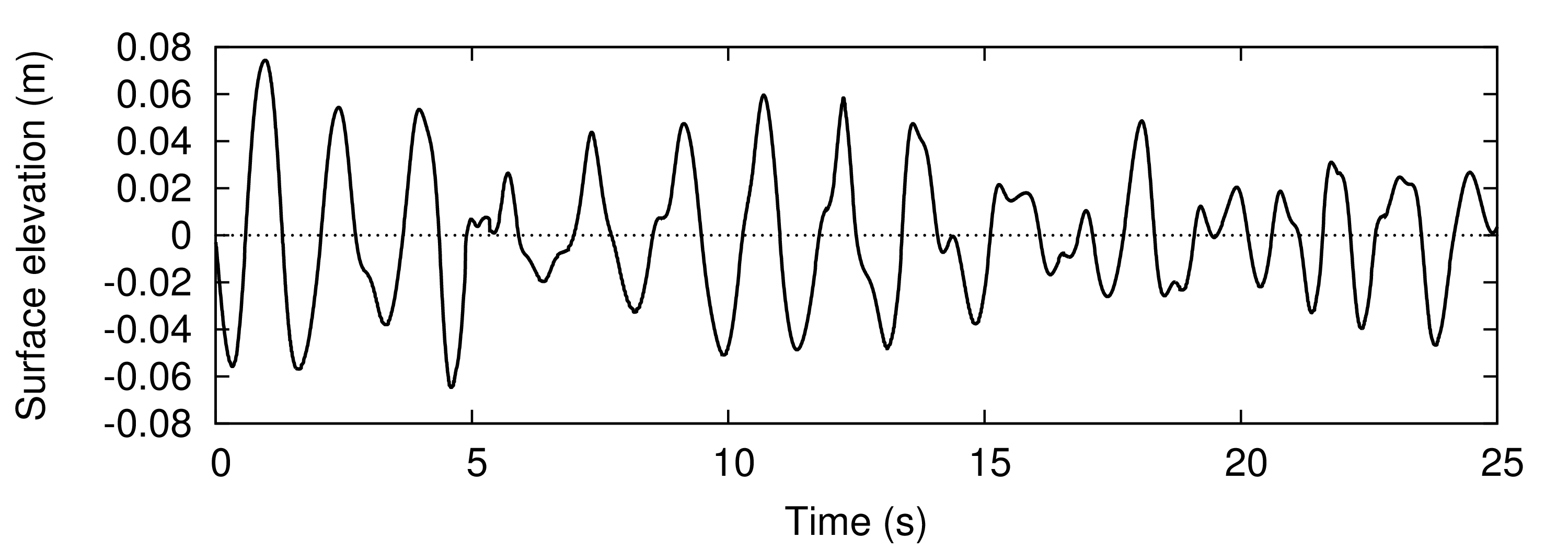}
\caption{Surface elevation over time recorded directly before the wave-maker} \label{FIGirr1}
\end{figure}

\begin{figure}[H]
\includegraphics[width=\factor\linewidth]{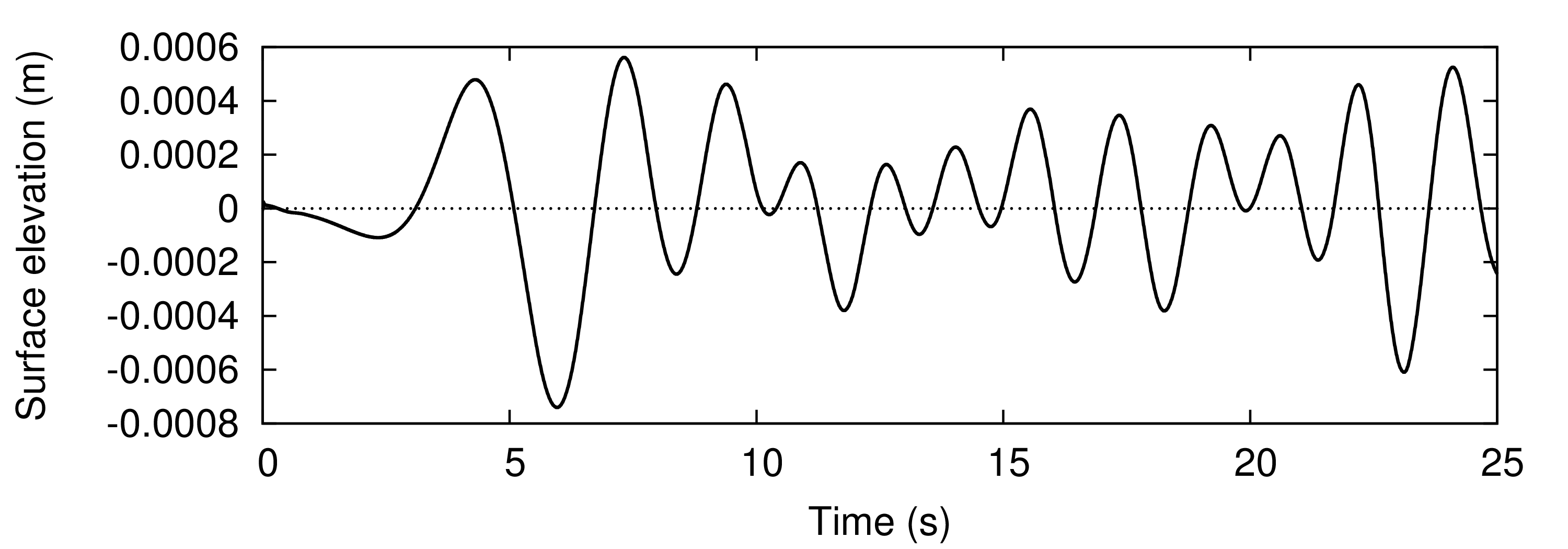}
\caption{Surface elevation over time recorded in close vicinity to the boundary, to which the damping layer is attached} \label{FIGirr2}
\end{figure}

\begin{figure}[H]
\includegraphics[width=\factor\linewidth]{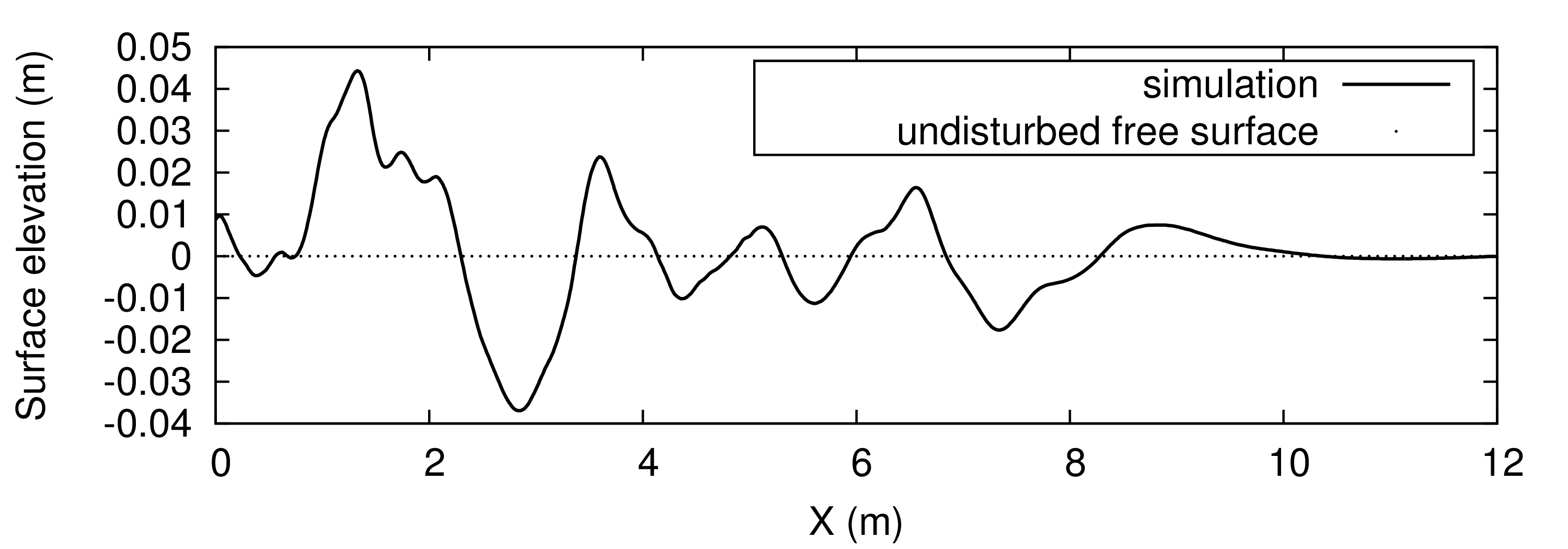}
\caption{Surface elevation in the whole domain at $ t \approx 15.6 \, \mathrm{s}$} \label{FIGirr3}
\end{figure}

The choice of $x_{\rm d} = 2 \lambda_{\rm peak} $, with wavelength $ \lambda_{\rm peak} $ corresponding to the peak wave frequency, seems adequate for such cases. No visible disturbance effects were noted in the simulation.
The wave heights over time are shown recorded close to the wave-maker  in Fig. \ref{FIGirr1} and recorded close to the boundary to which the damping zone is attached  in Fig. \ref{FIGirr2}. This shows that the average wave height is reduced by roughly two orders of magnitude. 
From the surface elevation in the tank in Fig. \ref{FIGirr3} no undesired wave reflections can be noticed. Although a more detailed study of the error of this approximation regarding different parameters of the spectrum was not possible in the scope of this study, the proposed approach seems to work reasonably well for practical purposes. Further research in this respect is recommended to reduce uncertainties regarding the reflections.

\section{Application of Scaling Law to Ship Resistance Prediction}
\label{kcs}
\begin{figure}[H]
\includegraphics[width=\factor\linewidth]{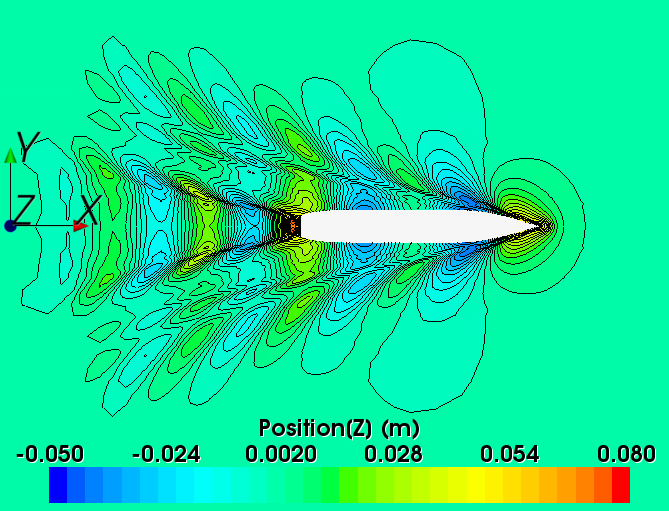}
\caption{Wave profile for model scale ship with $f_{1} = 22.5 \, \mathrm{s^{-1}}$ at $ t = 90 \, \mathrm{s} $} \label{FIGkcselev}
\end{figure}

This section compares results from model and full scale resistance computations in 3D of the Kriso Container Ship (KCS) at Froude number $ 0.26  $.
The simulations in this section are based on the simulations reported in detail in Enger et al. (2010).  For detailed information on the setup and discretization, the reader referred to Enger et al. (2010). In the following, only a brief overview of the setup and differences to the original simulations are given for the sake of brevity. The present model scale simulation differs only in the damping setup (and slight modifications of the used grid) from the fine grid simulation in Enger et al. (2010).  The KCS is fixed in its floating position at zero speed. To simulate the hull being towed at speed $U$, this velocity is applied at the inlet domain boundaries and  the hydrostatic pressure of the undisturbed water surface is applied at the outlet boundary behind the ship. The domain is initialized with a flat water surface and flow velocity $U$ for all cells. As time accuracy is not in the focus here, first-order implicit Euler scheme is used for time integration. Apart from the use of the $ k $-$ \epsilon $ turbulence model by Launder and Spalding (1974), the computational setup is similar to the one used in the rest of this work. The computational grid consists of roughly $ 3 $ million cells. For comparability, we compare only the pressure components of the drag and vertical forces, which are obtained by integrating the $x$-component for the  drag and $z$-component for the vertical component  of the pressure forces over the ship hull. The simulation starts at $ 0\, \mathrm{s} $ and is stopped at $ t_{\mathrm{max}} = 90\, \mathrm{s} $ simulation time.
The Kelvin wake of the ship can be decomposed into a transversal and a divergent wave component. The wave damping setup is based on the ship-evoked transversal wave (wavelength $\lambda_{\mathrm{t}} \approx 3.1 \, \mathrm{m}$), the phase velocity of which equals the service speed $U$. Wave damping according to Eqs. (\ref{damp1}) and (\ref{damp2}) has been applied to inlet, side and outlet boundaries with parameters $ x_{\mathrm{d}} = 2.3 \lambda_{\mathrm{t}}$, $ f_{1} = 22.5 \, \mathrm{s^{-1}}$, $ f_{2} = 0 $, and $ n = 2 $.
This setup provided satisfactory convergence of drag and vertical forces in model scale. The wake pattern for the finished simulation is shown in Fig. \ref{FIGkcselev} and the results are in agreement with the findings from Enger et al. (2010). The obtained resistance coefficient $C_{\mathrm{T,sim}} =3.533 \cdot 10^{-3} $ compares well with the experimental data ($C_{\mathrm{T,exp}} = 3.557 \cdot 10^{-3}$, $ 0.68\% $ difference to $ C_{\mathrm{T,sim}} $) by Kim et al. (2001) and to the simulation results by Enger et al. (2001)  ($C_{\mathrm{T,Enger}} = 3.561 \cdot 10^{-3}$, $ 0.11\% $ difference to $ C_{\mathrm{T,exp}} $).
Additionally, full scale simulations are performed with Froude similarity. The scaled velocity and ship dimensions are shown in Table \ref{TABkcs}. The grid is similar to the one for the model scale simulation except  scaled with factor $ 31.6 $. Assuming similar damping can be obtained with the presented scaling laws, $x_{\mathrm{d}}$ was scaled according to  Eq. \ref{xdscale}  by factor  $ 31.6 $ as well. Otherwise the setup corresponds to the one from the  model scale simulation. 
\begin{table}[H]
\caption{KCS parameters}\label{TABkcs}
\begin{center}
\tabcolsep=0.25cm
\begin{tabular}{c| c c}
\hline
scale & waterline length $ L\ (\mathrm{m})$ & service speed $U\ (\mathrm{m}/\mathrm{s})$\\ [0.5ex]
\hline
model & $7.357$ & $2.196$  \\ 
\hline 
full & $232.5$ & $12.347$ \\ 
\hline 
\end{tabular} 
\end{center}
\end{table}

\begin{figure}[H]
\includegraphics[width=\factor\linewidth]{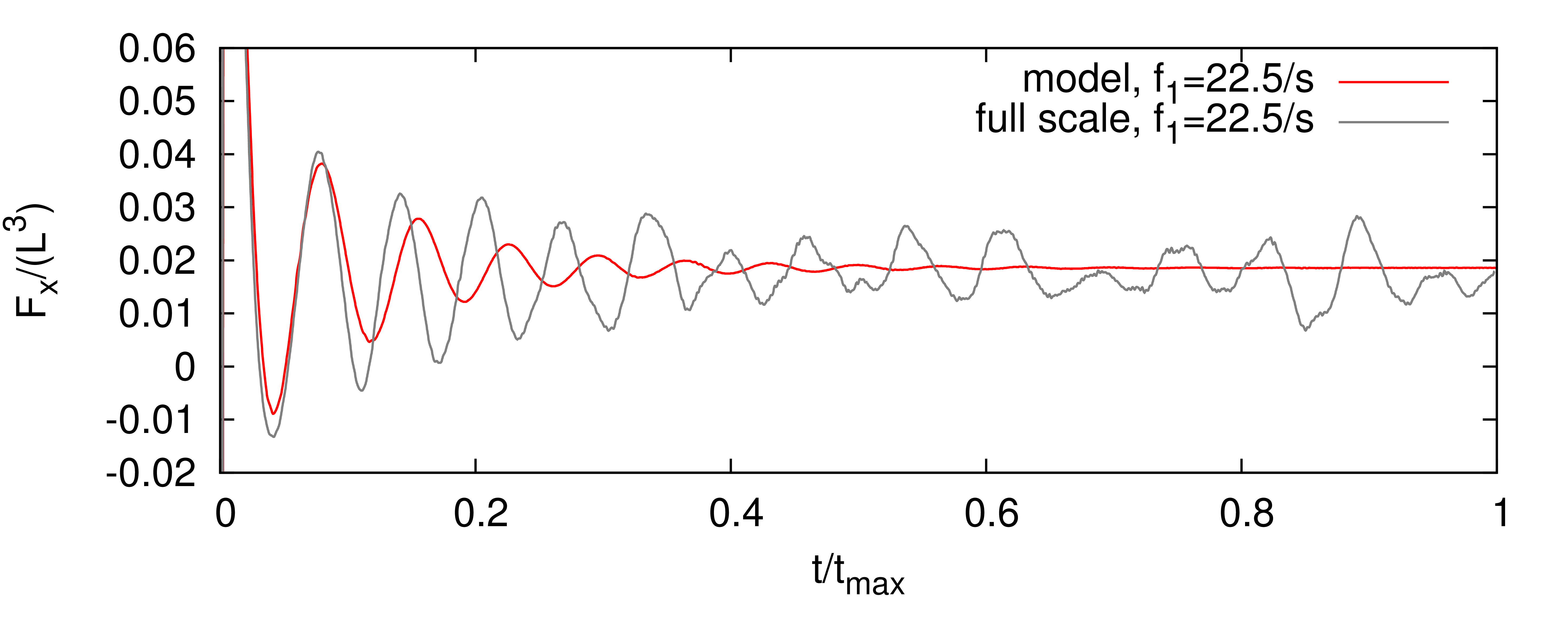}\\
\includegraphics[width=\factor\linewidth]{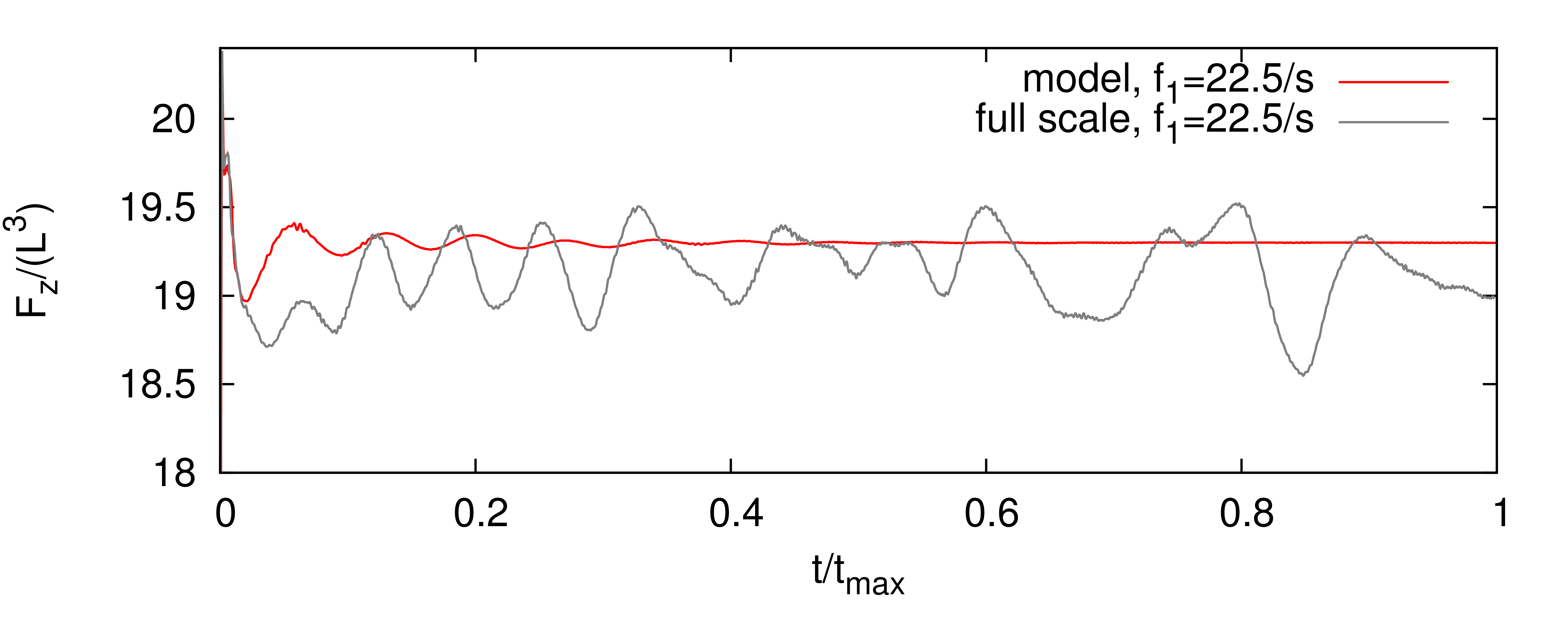}
\caption{Drag (top image) and vertical (bottom image) pressure forces on ship over time for model (red) and full scale (grey); no scaling of $ f_{1}  $, thus $ f_{1} = 22.5 \, \mathrm{s^{-1}} $ is the same in both simulations} \label{FIGdrag1}
\end{figure}

Figure \ref{FIGdrag1} shows drag and vertical forces over time when both model and full scale simulations are run with the same value for damping coefficient $ f_{1} $. In contrast to the model scale forces, the full scale forces oscillate in a complicated fashion. Therefore without a proper scaling of $ f_{1} $, no converged solution can be obtained for the full scale case.

\begin{figure}[H]
\includegraphics[width=\factor\linewidth]{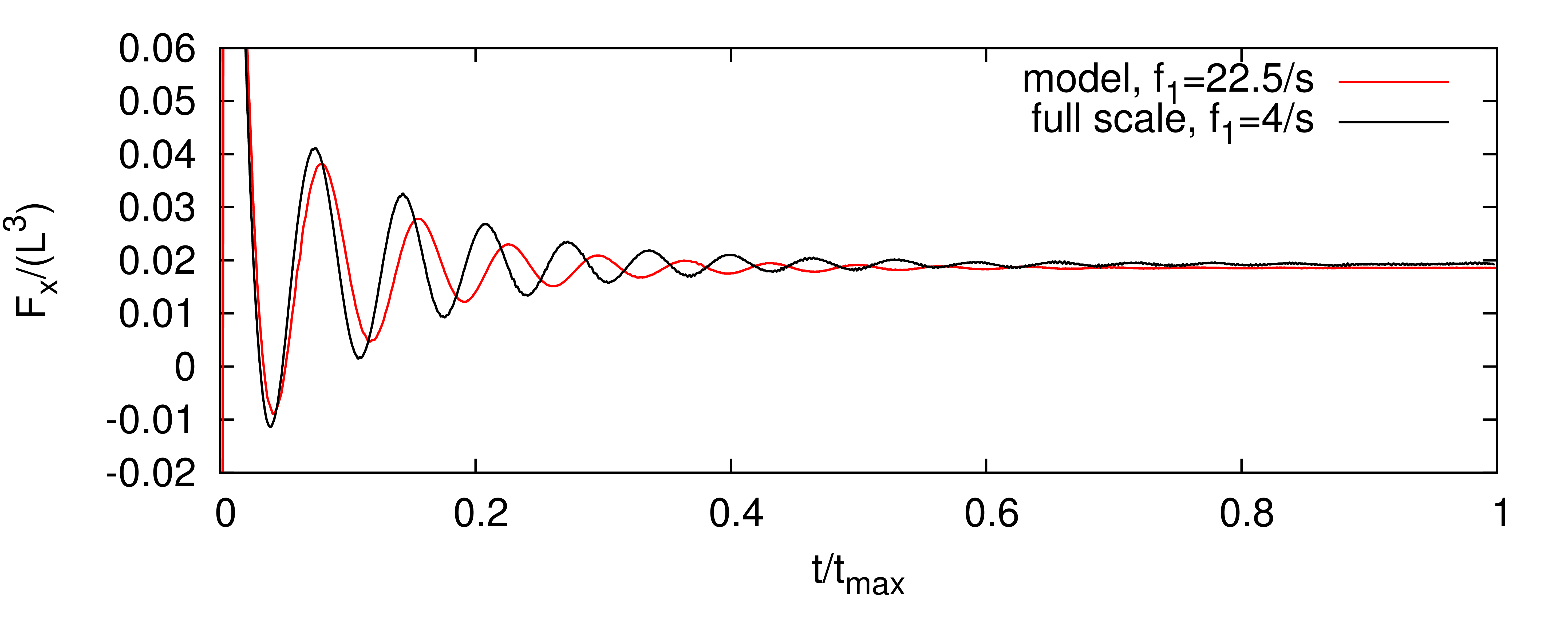}
\includegraphics[width=\factor\linewidth]{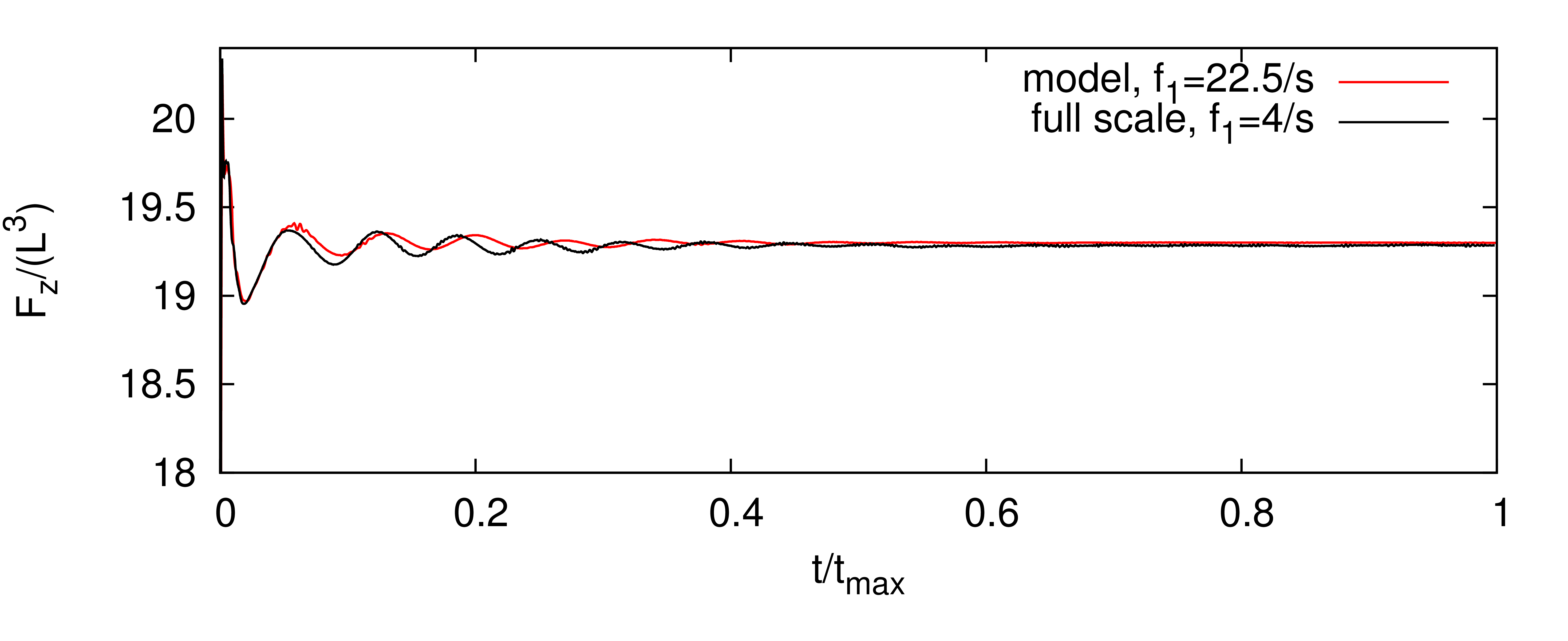}
\caption{Drag (top image) and vertical (bottom image) pressure forces on ship over time for model and full scale; the damping setup for the full scale simulation is obtained by scaling the model scale setup according to Eqs. \ref{xdscale} and \ref{omegascale}} \label{FIGdrag3}
\end{figure}

Finally, the full scale simulation is rerun with  $ f_{1}  $  scaled according to  Eq. \ref{omegascale} to obtain similar damping as in the model scale simulation with $ f_{1} =22.5 \, \mathrm{s^{-1}} $. The correctly scaled value for the full scale simulation is thus  $f_{1,\, \mathrm{full}} = f_{1,\, \mathrm{model}} \cdot \omega_{\mathrm{full}}/\omega_{\mathrm{model}} = 22.5 \, \mathrm{s^{-1}}\cdot 1/\sqrt{31.6} \approx 4 \, \mathrm{s^{-1}}$. The resulting drag and vertical forces in Fig. \ref{FIGdrag3}  show that indeed similar damping is obtained, since in both cases the forces converge in a qualitatively similar fashion. Note that a perfect match of the curves in Fig. \ref{FIGdrag3}  is not expected, since Froude-similarity is given, but not Reynolds-similarity. Thus although the pressure components of the forces on the ship will converge to the same values (if scaled by $L^{3}$), the way they converge (amplitude and frequency of the oscillation) may not be exactly similar, since this depends on the solution of all equations.
However, the tendency of the convergence will be qualitatively similar as shown in the plots.

\section{Discussion and Conclusion}
\label{concl}
In order to obtain reliable wave damping with damping layer approaches, the damping coefficients must be adjusted according to the wave parameters, as shown in Sects. \ref{l2f1}, \ref{l2f2}, \ref{Scalel2f1} and \ref{Scalel2f2}. It is described in Sect. \ref{assdamp} that the procedure to quantify the damping quality is quite effortful, and thus it is seldom carried out in practice.  This underlines the importance of the present findings for practical applications, since unless the damping quality can be reliably set to ensure that the influence of undesired wave reflections are small enough to be neglected, a large uncertainty will remain in the simulation results.
The optimum values for damping coefficients $f_{1}$ and $f_{2}$ can be assumed not to depend on  computational grid, wave steepness and thickness $x_{\rm d}$ of the damping layer as shown in Sects. \ref{gridinf}, \ref{lengthxd} and \ref{heightinfl}.
As shown in Sect. \ref{lengthxd}, the damping layer thickness has the strongest influence on the damping quality. If it increases, the range of  waves that will be damped satisfactorily broadens and the reflection coefficient for the optimum setup shrinks; unfortunately, the computational effort increases at the same time as well, thus optimizing the damping setup is important. 
In contrast, Sect. \ref{heightinfl} shows that the wave steepness has a smaller effect on the damping, with the tendency towards better damping for smaller wave steepness. For sufficiently fine discretizations, Sect. \ref{gridinf} shows that the damping can be considered not affected by the grid.
A practical approach for efficient damping of irregular waves has been presented in Sect. \ref{irrwav}. 
The scaling laws in Sect. \ref{Lf1} and recommendations given in Sects. \ref{Scalel2f1} and \ref{Scalel2f2}  provide a reliable way to set up optimum wave damping for any regular  wave.
Moreover, similarity of the wave damping can be guaranteed in model- and full-scale simulations as shown in Sects. \ref{Scalel2f1}, \ref{Scalel2f2} and \ref{kcs}. The findings can easily be applied to any implementation of wave damping which accords to Sect. \ref{general}.








\end{document}